\documentclass[aps,prl,twocolumn,showpacs,superscriptaddress]{revtex4-1}
\usepackage{graphicx,color,transparent}
\usepackage{psfrag}
\usepackage{dcolumn}
\usepackage{amsmath}
\usepackage{amssymb}
\usepackage{latexsym}
\usepackage{tikz}
\usepackage[plain]{fancyref}
\definecolor{mypurple}{RGB}{153,61,113}
\definecolor{myblue}{RGB}{63,61,153}
\definecolor{myokker}{RGB}{153,140,61}
\definecolor{mygreen}{RGB}{61,153,86}
\definecolor{mymarine}{RGB}{61,90,153}
\definecolor{mycyan}{RGB}{0,255,255}

\DeclareRobustCommand{\bdiamond}{%
  \mathbin{\text{\scalebox{.75}{\rotatebox[origin=c]{45}{\large
          $\Box$}}}}%
} \DeclareRobustCommand{\fdiamond}{%
  \mathbin{\text{\scalebox{.75}{\rotatebox[origin=c]{45}{\large
          $\blacksquare$}}}}%
}

\begin{document}

\def\bea{\begin{eqnarray}}
\def\eea{\end{eqnarray}}
\def\beq{\begin{equation}}
\def\eeq{\end{equation}}
\def\f{\frac}
\def\k{\kappa}
\def\e{\epsilon}
\def\ve{\varepsilon}
\def\be{\beta}
\def\D{\Delta}
\def\h{\theta}
\def\t{\tau}
\def\a{\alpha}

\def\rv{{\bf r}}
\def\vv{{\bf v}}
\def\jv{{\bf j}}
\def\kv{{\bf k}}
\def\Gv{{\bf G}}

\def\eff{{\rm eff}}
\def\mf{{\rm mf}}
\def\cDa{{\cal D}[X]}
\def\cD{{\cal D}[x]}
\def\cL{{\cal L}}
\def\cLo{{\cal L}_0}
\def\cLa{{\cal L}_1}

\def\Re{{\rm Re}}
\def\sj{\sum_{j=1}^2}
\def\rk{\rho^{ (k) }}
\def\rek{\rho^{ (1) }}
\def\cek{C^{ (1) }}
\def\rz{\rho^{ (0) }}
\def\rt{\rho^{ (2) }}
\def\rtb{\bar \rho^{ (2) }}
\def\trk{\tilde\rho^{ (k) }}
\def\trek{\tilde\rho^{ (1) }}
\def\trz{\tilde\rho^{ (0) }}
\def\trt{\tilde\rho^{ (2) }}
\def\tD{\tilde {D}}

\def\s{\sigma}
\def\kb{k_B}
\def\la{\langle}
\def\ra{\rangle}
\def\nn{\nonumber}
\def\up{\uparrow}
\def\dn{\downarrow}
\def\S{\Sigma}
\def\dg{\dagger}
\def\d{\delta}
\def\p{\partial}
\def\l{\lambda}
\def\L{\Lambda}
\def\G{\Gamma}
\def\o{\Omega}
\def\w{\omega}
\def\g{\gamma}

\def\noi{\noindent}
\def\a{\alpha}
\def\d{\delta}
\def\p{\partial} 

\def\la{\langle}
\def\ra{\rangle}
\def\e{\epsilon}
\def\n{\eta}
\def\g{\gamma}
\def\break#1{\pagebreak \vspace*{#1}}
\def\hf{\frac{1}{2}}

\title{Stochastic ratcheting of two dimensional colloids : Directed
  current and dynamical transitions} \author{Dipanjan Chakraborty}
\affiliation{
  Indian Institute of Science Education and Research, Mohali,
  Punjab-140306, India.  }  
\email{chakraborty@iisermohali.ac.in}
\author{Debasish Chaudhuri} 
\affiliation{Indian Institute of
  Technology Hyderabad, Yeddumailaram 502205, Andhra Pradesh, India }
\email{debc@iith.ac.in}

\date{\today}
\begin{abstract}
  We present results of molecular dynamics simulations for
  two-dimensional repulsively interacting colloids driven by a one
  dimensional asymmetric and commensurate ratchet potential, switching
  on and off stochastically. This drives a time-averaged 
  directed current of colloids, exhibiting resonance with change in ratcheting
  frequency, where the resonance frequency itself depends non-monotonically on
  density. Using scaling arguments, we obtain analytic results that show
  good agreement with numerical simulations.  With increasing ratcheting frequency,
  we find {\em non-equilibrium re-entrant transitions} between solid and
  modulated liquid phases.
\end{abstract}

 \maketitle

A flashing ratchet refers to a time-averaged directed motion of
Brownian particles under the influence of an spatially periodic and
asymmetric potential, with the potential height varying with time, either deterministically or stochastically~\cite{Julicher1997a,
  Reimann2002,Astumian2002,Hanggi2009}. Stochastic ratcheting has been
 studied extensively, in the context of active dynamics of molecular
motors ~\cite{Prost1994,Julicher1995,Julicher1997,Astumian1997}, dynamics of
colloidal dispersion in electrical~\cite{Rousselet1994,Leibler1994,Marquet2002},
magnetic~\cite{Tierno2010,Tierno2012} or optical
drive~\cite{Faucheux1995, Lopez2008}, as a mechanism of particle
segregation~\cite{Chou1999,Kettner2000,Matthias2003},
transport of cold atoms in optical
lattice~\cite{Mennerat-Robilliard1999}, and in the motion of flux
quanta~\cite{Lee1999,Olson2001}.  While a large body of work has been
concentrated on the ratcheting of individual particles, fewer studies
focused on the effects of interaction
~\cite{Derenyi1995,Derenyi1996,Reimann1999,Rapaport2002,Aghababaie1999}.
Recent studies of two dimensional (2D) paramagnetic particles under
one dimensional (1D) magnetic ratchets observed relation between
overall dynamics and local  particle coordination numbers~\cite{Tierno2012}. 

\begin{figure}[!t]
\begin{center}
\includegraphics[width=8 cm] {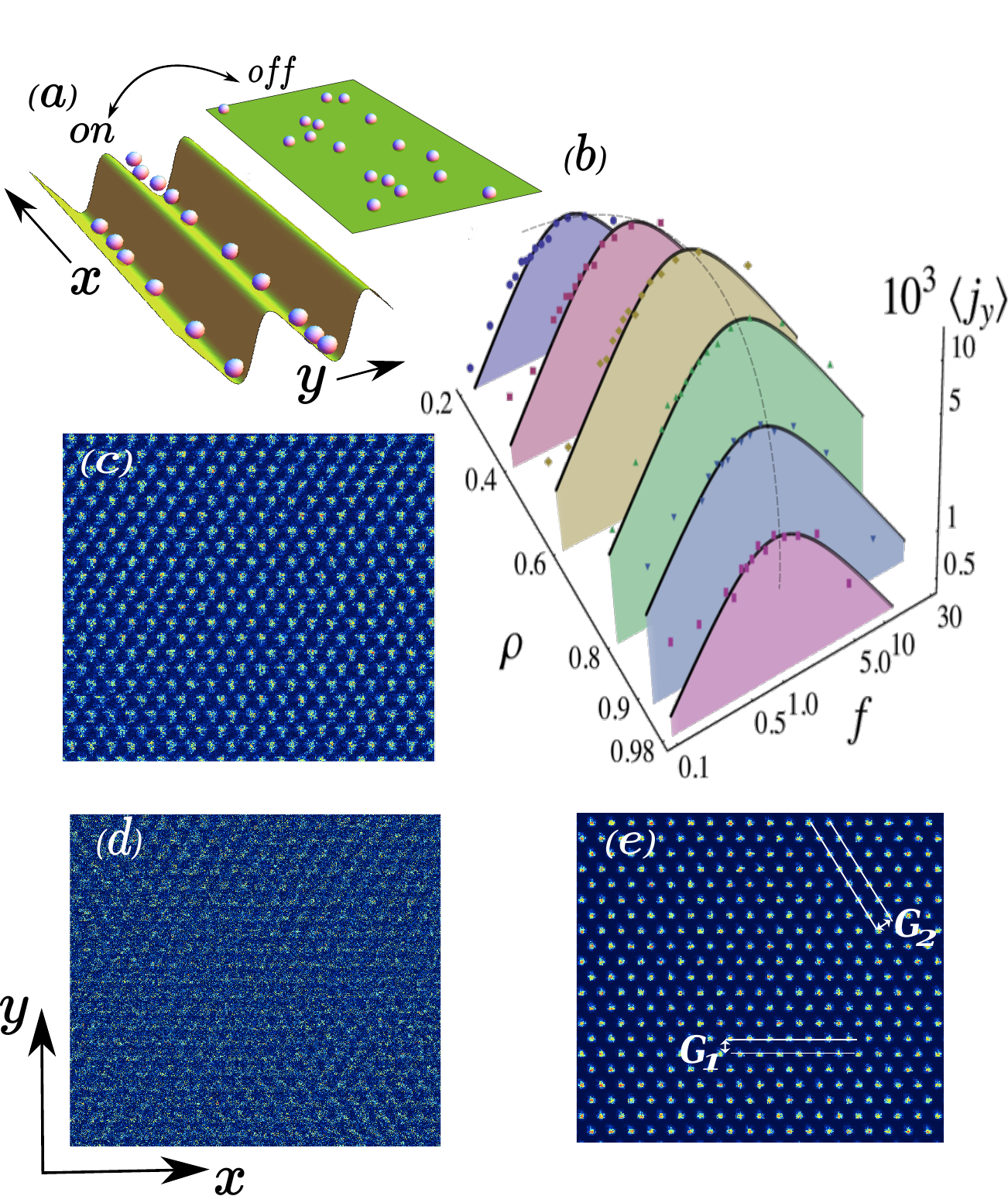}  
\caption{(Color online) 
($a$) Schematic of 2d colloids in 1D asymmetric ratchet
  potential periodic in $y$-  and constant in $x$- direction.  The
  arrow between the (green) corrugated and flat surfaces denote   
  switching of external potential between {\em on} and {\em off} state with rate $f$.
$(b)$~Time-averaged directed current along $y$-direction $\la j_y\ra$ as a function of frequency $f$, and density $\rho$.
The dashed line indicates variation of maximal current with density.
  ($c$) -- ($e$): Superimposed positions of $10^3$ uncorrelated
  configurations, from center of mass coordinates, at
  a density $\rho  = 1.0$ with ratcheting frequencies
  ($c$)~$f=0.11$, ($d$)~$f=1.67$
  and ($e$)~$f=10$. 
  The color code denotes local density of points from red/light (high)
  to blue/dark (low). 
 The reciprocal lattice vectors ${\bf G}_{1,2}$ and the corresponding lattice planes are indicated in ($e$).
  }
\label{fig:jdcq}
\end{center}
\end{figure}

In colloidal suspensions, ratchet-like directed motion of particles
have been achieved using suitable laser potentials~\cite{Faucheux1995,Lopez2008}.
Confinement and laser trapping in colloids, on the other hand, is known to give rise to interesting mechanical
properties and phase
transitions~\cite{Henseler2006,Mangold2003,Chaudhuri2004,U.Siems2012,Chowdhury1985,Wei1998}.
Coupling 2D interacting colloids to a 1D time-independent spatially
periodic potential with periodicity commensurate with the mean
particle separation, leads to the phenomena of laser induced freezing
(LIF) and re-entrant melting with increase in the potential strength.
This was demonstrated in experiments using standing wave pattern of
interfering laser beams~\cite{Chowdhury1985,Wei1998}, and was
understood in terms of a dislocation unbinding
theory~\cite{Frey1999,Chaudhuri2006}.

We consider transport of a 2D system of  particles interacting via soft-core repulsion and driven by
an 1D asymmetric flashing ratchet, using molecular dynamics (MD)
simulations in the presence of a Langevin heat bath.  The ratcheting potential
breaks time-reversal symmetry and generates an average directed
current along the direction of ratcheting (\Fref{fig:jdcq}(a)).  We
choose a periodicity of the potential commensurate with the inter
particle separation. At switching frequencies much faster than the
intrinsic relaxation times, the time scale required for particles to relax over a single valley of the external ratchet potential, 
the system experiences a time-averaged
effective periodic potential, which in the limit of weak asymmetry is
expected to lead to a situation similar to that of LIF.
However, at intermediate switching frequencies the system is driven out of equilibrium and carries an averaged directed current.
We present the transport properties, and relation between structure and transport.

In the 2D ratchet system that we study, averaged directed current shows
non-monotonic variation with density and ratcheting frequency, with
the maximal current achieved at their intermediate values. 
The behavior differs significantly in detailed functional dependence from 1D ratchet.  
Using scaling arguments, we derive
expressions for the directed current which fully capture the
simulation results.  Our study on 2D ratchet reveals two fascinating
properties which are unlike 1D ratchet: 
(i)~with increasing ratcheting frequency we find {\em re-entrant non-equilibrium phase transitions} between solid and modulated liquid phases,
 as averaged directed current shows non-monotonic variation, 
(ii)~crossover from ballistic to diffusive transport with density, captured by a non-monotonic density-dependence of resonance frequency.
Our predictions are amenable to verification  in experiments on,
e.g., sterically stabilized colloids driven by suitably tunable optical or magnetic
ratchets~\cite{Tierno2012,Faucheux1995}.

{\em Model:}
As a model colloid, we consider a system of purely repulsive
particles interacting via a shifted and truncated  soft-core potential 
$\be U(r)=(\sigma/r)^{12}-2^{-12}$ with a cutoff distance
$r_c=2\sigma$, so that $\be U(r)=0$ for $r > r_c$. Here $\kb T=1/\be$
and $\s$ set the energy and length scales, respectively.
The asymmetric ratchet potential $U_{\rm ext}(y,t)=V_0(t)
\left[\sin\left( 2 \pi y/\l \right) +\alpha \sin\left(4\pi y/\l
  \right) \right]$, where $V_0(t)$ switches between $U_0$ and $0$ with
a switching rate $f$, which we also refer to as frequency.
We use  the asymmetry parameter $\a=0.2$ (see~\Fref{fig:jdcq}(a)).
In all our simulations we set $\be U_0=1$.  The external potential is
kept commensurate to the density of the particles, such that $\l=a_y$,
with the separation between consecutive lattice planes $a_y=\sqrt{3}
a/2$ in a triangular lattice at a density $\rho=2/\sqrt 3
a^2$. 
MD simulations are performed using the standard
leap-frog algorithm~\cite{Frenkel2002} with a time step $\d
t=0.001 \t$, where $\tau=\sigma\sqrt{m/\kb T}$ is the characteristic
time scale.  We choose the mass of the particles $m=1$, and set the
temperature $T=1.0\, \e/\kb$ by using a Langevin
thermostat~\cite{Grest1986} with an isotropic friction $\g = 1/\t$.
At each time step, a trial move to perform switching of the external potential strength between $0$ and $U_0$
is performed, and accepted with probability $f\d t$.
We used $N=4096$ particles in our simulations. 

\begin{figure}[t]
\hspace{-0.7cm}
\includegraphics[width=7cm]{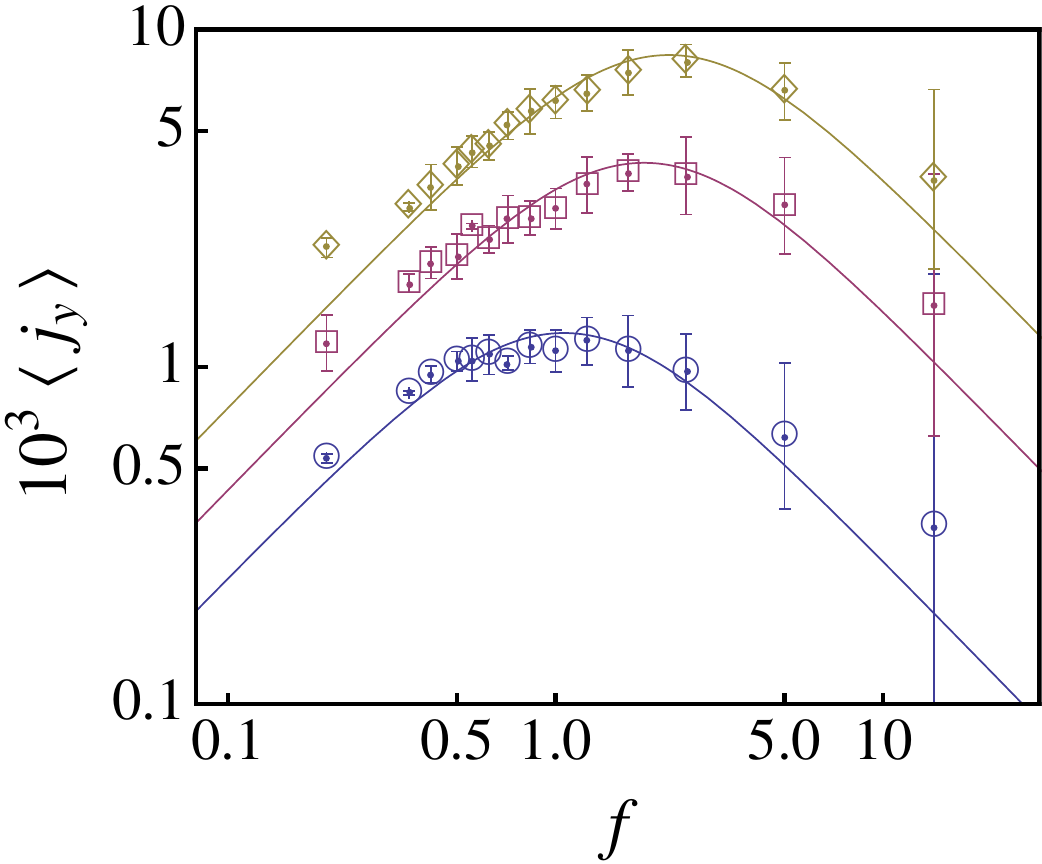}
\caption{(Color Online) Average directed current as a function
  of switching rate $f$, at particle densities,
  $\rho=0.1$({\color{myblue} {\Large $\circ$}}),
  $0.5$~({\color{mypurple}\footnotesize $\square$}) and
  $1.0$~({\color{myokker}{\large $\diamond$}}).
  The solid lines show fit 
  to Eq.~\ref{eq:flux_scaling_form}. }
  \label{fig:flux_freq}
\end{figure}

The soft-core particles, in the absence of external potential, freezes
at a density $\rho^\ast \approx 1.01$ (see Fig.1($a$) in the
Supplemental Material~\cite{supmat}).  The limit of $\a=0$ and
$V_0(t)=U_0$ corresponds to the equilibrium situation
of laser induced freezing~\cite{Frey1999}. At a density close to the
liquid-solid transition, the system freezes into a triangular lattice
solid (LIF) which remelts into a density modulated liquid with
increasing $U_0$~\cite{Frey1999,Wei1998}.  In soft-core particles, the
LIF with $\be U_0=1$ occurs at $\rho = 0.95$~\cite{Chaudhuri2006}.
Similar freezing transition at this density is observed for a weakly
asymmetric ratchet ($\a=0.2$) of strength $\be U_0=1$ in the limit of
high switching frequency, much faster than the typical relaxation
time, such that the colloids experience an effective periodic
potential (see Fig.2($d$) in the Supplemental
Material~\cite{supmat}). In the other limit of extremely slow
switching, the system comes to {\em quasi}-equilibrium with the
instantaneous strength of external potential, and one obtains a slow
variation between a modulated liquid and a solid phase.  The most
interesting dynamics takes place at intermediate frequencies.  
The ratchet- driven averaged directed current shows resonance with frequency, 
and non-monotonic variation with density (\Fref{fig:jdcq}($b$)).
At suitable densities, the system shows dynamical re-entrant transition from a soft solid to
modulated liquid to solid with increase in ratcheting
frequency~(\Fref{fig:jdcq}($c$)--($e$) and \Fref{fig:solid_op_freq}).  

{\em Transport properties:} The steady state dynamics is characterized
in terms of a space and time-averaged directed current of particles
flowing along the direction of ratcheting
\begin{equation}
 \la j_y \ra = \f{1}{\t_m}
\f{1}{L_x L_y} \int^{\t_m} dt \int^{L_x} dx \int^{L_y} dy
\,\,j_y(x,y,t)
\label{eq:jy}
\end{equation} 
where the time averaging is done over $\t_m= n t_p$, with $t_p=1/f$
and $n$ denotes a large number of switching, chosen to be $200$ in all
our simulations.

\begin{figure}[t]
  \centering
  \includegraphics[width=7 cm]{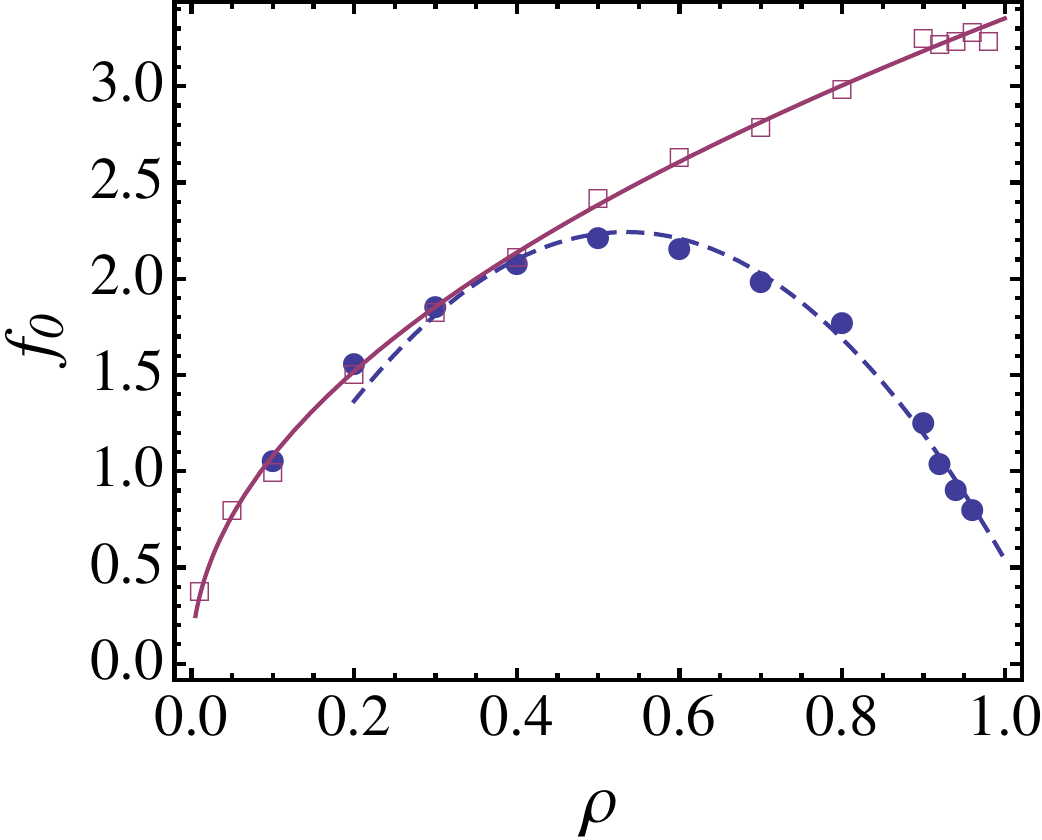}
  \caption{(Color Online) Resonance frequency $f_0$ as a function of
    the density $\rho$ for non-interacting particles (squares),
    and soft-core particles (circles). 
    The solid line shows ballistic form $f_0 \sim \sqrt{\rho}$, while the
    dashed line shows diffusive form $f_0 \sim \rho (1-\rho/\rho_c)$ with
    $\rho_c = 1.07$. }
  \label{fig:resonance_comparison}
\end{figure}

For small switching frequencies $f \ll \nu$, the inverse of intrinsic relaxation time, the system is close to 
thermodynamic equilibrium. The directed current increases as $\la j_y \ra \sim f$ starting from zero at $f=0$ in agreement with linear response~\cite{Ajdari1992,Prost1994,Luczka1997}. 
The frequency dependence at high switching rate was calculated earlier using asymptotic expansion
to give $\la j_y \ra \sim 1/f$~\cite{Luczka1997,Bao1998}. 
 
In our MD simulations of 2D system of soft-disks, we observe the same behavior, viz., $\la j_y \ra \sim f$ at low frequency, 
and $\la j_y \ra \sim 1/f$ at very high ratcheting frequencies~(\Fref{fig:flux_freq}).  The asymptotic behavior may be captured by
the interpolation formula $g(\nu,f) = \nu f/(\nu^2 +f^2)$. 
We use a simple ansatz $\la j_y \ra = \k g(\nu,f)\rho v_0$ where $\k$ is a
dimensionless proportionality constant, and $\rho v_0$ has the
dimension of current with $v_0$ an intrinsic velocity. As we show below, the form of
$v_0$ and $\nu$ allow us to describe the whole density and frequency dependence of directed current.
The relation
\begin{equation} 
\la j_y \ra = \k \f{\nu f }{\nu^2 +  f^2} \, \rho v_0,
\label{eq:flux_scaling_form}
\end{equation}
shows good agreement with simulation results
(\Fref{fig:flux_freq}).  The above frequency dependence is obeyed
even if the ratcheting wavelength $\l$ is incommensurate with density
(see Fig.2($a$)-($c$) in the Supplemental Material~\cite{supmat}).  A
similar frequency dependence was recently found for a stochastic pump
model of one-dimensional system of interacting
particles~\cite{Chaudhuri2011}.  Fitting the MD simulation data of
\Fref{fig:flux_freq} to Eq.(\ref{eq:flux_scaling_form}) we find
the resonance frequencies $f=f_0=\nu$ which show a non-monotonic
variation with the mean density of colloids $\rho$ (\Fref{fig:resonance_comparison}).

The intrinsic relaxation frequency $\nu$, controlling the behavior of
time-averaged dynamics, may arise from a ballistic or diffusive
relaxation of the particles over the characteristic length scale $\l$.
We use ratcheting potential commensurate with the density such that $\l^2
\sim 1/\rho$ (for treatment using incommensurate potential, see Supplemental Material~\cite{supmat}). 
For under-damped motion, the ballistic time-scale $\t_b$
for a particle to traverse the potential valley is obtained from the
kinematic relation $\l \sim (U_0/\l) \t_b^2$, that leads to $\t_b \sim
(\rho U_0)^{-1/2}$.
On the other hand,
the relaxation time in the over-damped diffusive regime
is given by $\t_D = \l^2/D \sim (D \rho)^{-1}$.
The self diffusion constant $D$ decreases with density for
two-dimensional repulsively interacting particles as $D =
D_0(1-\rho/\rho_c)$~\cite{Lahtinen2001,Falck2004} (see Fig.1($b$) in the Supplemental
Material~\cite{supmat}).

\begin{figure}[t]
  \centering
\includegraphics[width=7 cm]{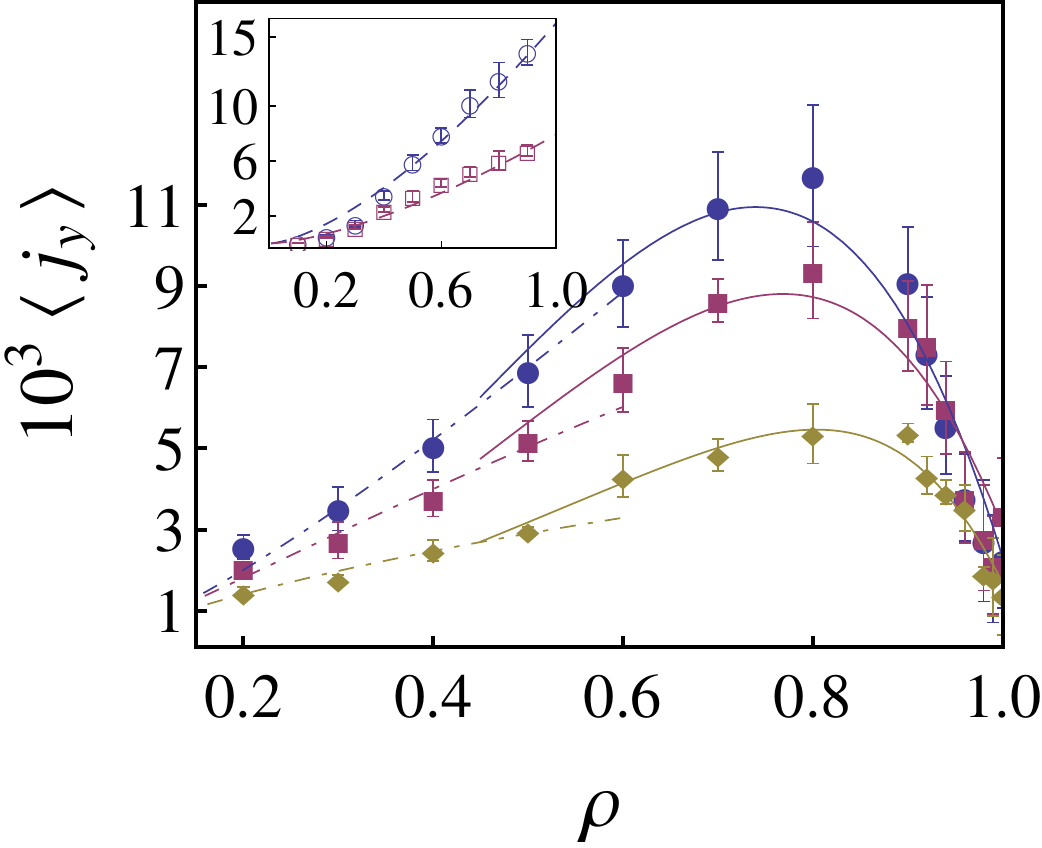}
\caption{(Color Online) 
Average particle flux $\la j_y \ra$ as a
  function of density $\rho$, for soft-core particles 
  at ratcheting frequencies
  $f = 1.43 $ ~({\color{myblue} {\large $\bullet$}}), $0.71 $
  ~({\color{mypurple}\footnotesize $\blacksquare$}), $
  0.36$~({\color{myokker} {$\fdiamond$}}).The dot-dashed
  lines are fit to Eq.(\ref{eq:ballistic}) in the regime $\rho < 0.5$
 with fitting parameter $\k = 0.04, 0.03,
  0.02$ for the three data sets respectively.  The solid lines are fit to 
  Eq.(\ref{eq:diffusive}) 
  in the regime $\rho \geq 0.5$
  with 
  fitting parameters $\k = 0.12, 0.05, 0.02$ and $\rho_c = 1.03,\,1.05,\, 1.03$. 
  Inset: The same quantity for free particles at two different 
  ratcheting frequencies $f  = 0.71$~({\color{myblue} {\Large
      $\circ$}}) and $ 0.36$~({\color{mypurple}\footnotesize
    $\square$}).  The dashed lines are fit to 
  Eq.(\ref{eq:ballistic}).
 }
  \label{fig:fluxplot_dens_linscale}
\end{figure}

In the underdamped case, the velocity scale is set by $v^b_0 = \l/\t_b=U_0^{1/2}$. 
Using this and $\nu=1/\t_b$ in this regime, one finds
\begin{equation}
\label{eq:ballistic}
\la j_y \ra \simeq \k \f{f U_0}{\rho U_0 + f^2} \rho^{3/2}.
\end{equation}
The resonance frequency is then $f_0=(\rho U_0)^{1/2}$.
On the other hand, the velocity scale in the over-damped
regime may be obtained using the time-scale for free diffusion $1/(\rho
D_0)$ over mean inter-particle separation $\l$, $v^D_0 = D_0 \rho^{1/2}$.
Thus, using $\nu=1/\t_D$ the averaged directed current becomes
\begin{equation}
  \la j_y \ra \simeq \k \f{f D_0^2}{D_0^2 \rho^2 (1-\rho/\rho_c)^2 +
    f^2} \rho^{5/2}(1-\rho/\rho_c).
\label{eq:diffusive}
\end{equation}
The corresponding resonance frequency is $f_0=D_0 \rho (1-\rho/\rho_c)$.

Our simulations show that the resonance frequency, and therefore the
intrinsic relaxation frequency, follows ballistic behavior $f_0 \sim
\sqrt \rho$ at low densities (and for non-interacting particles), and
diffusive behavior $f_0 \sim \rho (1-\rho/\rho_c)$ at high
densities~(\Fref{fig:resonance_comparison}).
The dynamical behavior changes from ballistic to diffusive with
increase in density.  This may be understood in terms of what happens
to a directed current in the presence of direction randomizing
scattering events.  At low densities, the time- and space- averaged
motion of a test particle with small number of scattering events
remains ballistic on an average.
However, at large densities mean free path reduces, and
consequently, a large number of scattering events randomizes the
direction of motion leading to a predominantly diffusive dynamics.

\begin{figure}[t]
{\includegraphics[scale=0.45]{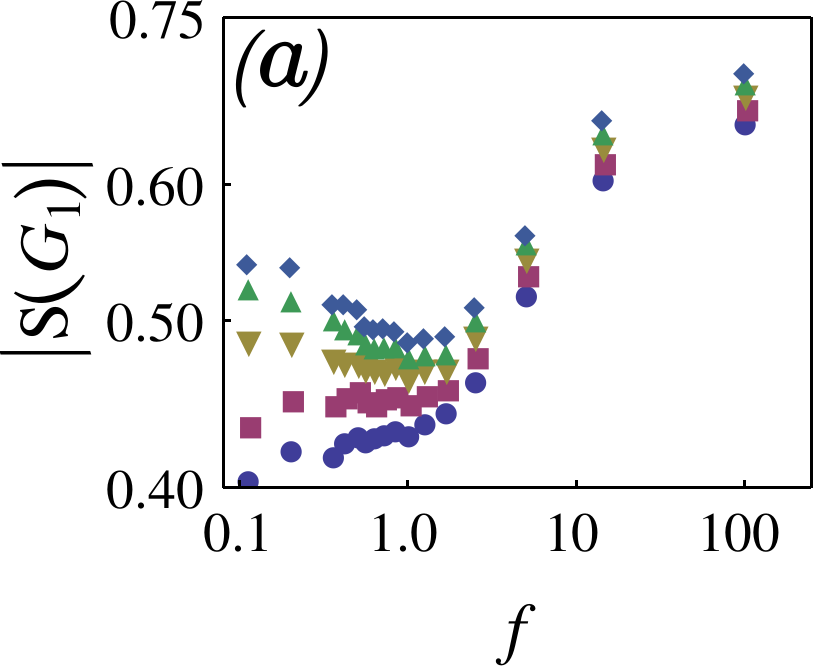}}
{\includegraphics[scale=0.45]{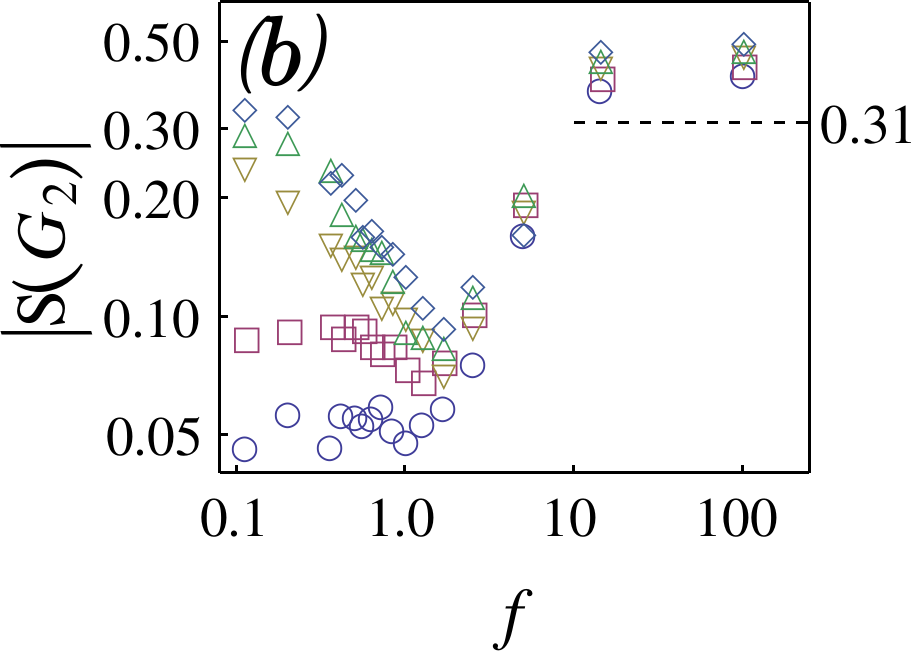}}
\caption{(Color Online) Amplitude of steady state structure factor, for the
  reciprocal lattice vectors $\mathbf{G}_1$ (a) and
  $\mathbf{G}_2$ (b), as a function of ratcheting frequency $f$
  for densities $\rho=0.98$~({\color{myblue}{\large$\bullet$},
    {\large$\circ$}}), $0.99$~({\color{mypurple}\footnotesize
    $\blacksquare$},{\color{mypurple}\footnotesize $\square$}), $1.00
  $~({\color{myokker}$\blacktriangledown$,$\triangledown$}),
  $1.01$~({\color{mygreen}$\blacktriangle$,$\triangle$}) and
  $1.02$~({\color{mymarine}{{\large $\fdiamond$},\large $\diamond$}}).}
  \label{fig:solid_op_freq}
\end{figure}

In \Fref{fig:fluxplot_dens_linscale} we show the density
dependence of the directed current at various switching
frequencies. The plots show non-monotonic variation, the low density
limit of which is fully captured by Eq.(\ref{eq:ballistic}), 
and the high density limit by
Eq.(\ref{eq:diffusive}). Near the density $\rho_c$, the
system gets into a {\em jammed} state where the directed current vanishes as
$\la j_y \ra \sim \rho^{5/2} (1-\rho/\rho_c)$.
Note that, the overall density- dependence that we find in 2D ratchet 
is quite unlike the $\la j \ra \sim \rho(1-\rho)$ behavior of directed 
current found in repulsively interacting 1D ratchet~\cite{Aghababaie1999}. 
%
The collective dynamics of the 2D ratchet can be further characterized in
terms of the density and ratcheting- frequency dependence of
longitudinal and transverse diffusivity $D_{x,y}(\rho,f)$
(see Fig.s 3 and 4 in the Supplemental Information~\cite{supmat}).

{\em Dynamical transitions:} The reduction of directed
current at high densities and subsequent jamming
is associated with freezing of the system into a triangular lattice
solid. 
Our MD simulations showed similar structural transitions are also associated with change in current 
as a function of  ratcheting frequency (\Fref{fig:jdcq}($b$)), a fully dynamical effect.
In \Fref{fig:jdcq}($c$)--($e$),  
we plot the superimposed positions of $10^3$ uncorrelated
configurations from the center of mass frame,  
for a system at a mean density $\rho=1.0$, and ratcheting frequencies $f
= 0.11,\, 1.67,\,10 $. 
This suggests frequency dependent re-entrant {\em transition} from a
triangular lattice solid ($f=0.11$), to density modulated liquid
($f=1.67$), to again a triangular lattice solid ($f=10$) order.  Note
from \Fref{fig:flux_freq} that the modulated liquid at
$\rho=1.0$ and $f=1.67$ corresponds to the resonance frequency in
directed current.

The interplay of structure and dynamics is further quantified with
the help of  time-averaged
steady state structure factor $S(\mathbf{G})=\la \frac{1}{N^2}
\sum_{i,j} \exp({-\mathbf{G}. (\mathbf{r}_i-\mathbf{r}_j)}) \ra $ with
reciprocal lattice vectors  $\mathbf{G}_1=(0,\pm 2\pi/a_y )$  
and $\mathbf{G}_2=(\pm 2 \pi/a, \pm 2 \pi / \sqrt{3}a)$  (see \Fref{fig:jdcq}($e$)). 
In \Fref{fig:solid_op_freq} we show the frequency dependence of  $|S(\mathbf{G_{1,2}})|$
at various densities.  The presence of ratcheting potential keeps
$|S(\mathbf{G_{1}})| > |S(\mathbf{G_{2}})| $
corresponding to stronger  density modulation in the $y$-direction.  
The non-monotonic variation of
$|S(\mathbf{G_{1}})|$ with frequency quantifies a reduction followed
by an increase in this density- modulation.  At very
high frequencies, the solid- order parameter $|S(\mathbf{G_{2}})| >
0.31$ for densities $\rho \gtrsim 0.96$, signifying
freezing into a triangular lattice structure  (see Fig.s 1($a$) and 2($d$) in Supplemental Material~\cite{supmat}), reminiscent of LIF
transition~\cite{Chaudhuri2006}.  The solid order parameter
$|S(\mathbf{G_{2}})|$ at densities $\rho \geq 1$ shows significant
non-monotonic variation with frequency, pointing to a dynamical
re-entrant {\em transition} from a solid to modulated liquid to solid
phase.  Thus ratcheting frequency provides a means to structural
control during transport, and may be utilized in experiments.

{\em Summary and outlook:}
Our study on a 2D system of soft-core particles under 1D ratchet drive, have shown
interesting relation between transport properties and structural phases.
Using scaling arguments we 
obtained the density and ratcheting frequency dependence of averaged directed current $\la j_y \ra$,
which fully captured the simulation results. The resonance frequency of $\la j_y \ra$  showed a 
curious cross-over from ballistic to diffusive behavior with increasing density, related to reduction of mean free path.
Within a range of densities, we found a {\em dynamical re-entrant transition} from solid- to modulated 
liquid- to solid- phase with increasing ratcheting frequency. The fact that ratcheting frequency provides a control over 
both the emergent directed current and structural phases, may have useful applications.

Our predictions may be verified in experiments on
repulsively interacting colloids confined within glass plates, e.g., using magnetic ratcheting~\cite{Tierno2012},
or optical ratcheting~\cite{Faucheux1995} in a suitably modified 2D laser trapping setup~\cite{Wei1998}. 
For example, polysterene beads have density $1.05 \,{\rm g/cm}^3$, 
i.e., a bead of diameter  $\s \approx 5\,\mu$m have mass $m \approx 6.9\times 10^{-14}\,$Kg. Given 
$\kb T = 4.2 \times 10^{-21}\,$Nm at room temperature, the unit of time $\t = \s \sqrt{m/\kb T} \approx 0.02 \,$s. 
Thus  the dimensionless frequency range of $f=0.1$ to $100$ studied here, corresponds to a range of 5\,Hz to 5\,KHz,
and the resonance at $f_0 \approx 1$ means a frequency of 50\, Hz.

\noindent{\em Acknowledgment:}
We thank Madan Rao for a valuable suggestion. Debasish Chaudhuri thanks Surajit Sengupta, Sriram Ramaswamy,  Narayanan
Menon, Swarnali Bandopadhyay for  useful discussions, Abhishek Chaudhuri, Bela M. Mulder for critical 
comments on the manuscript, and MPI-PKS Dresden for hosting him at various stages of this work.

\bibliographystyle{prsty}

\begin{thebibliography}{10}

\bibitem{Julicher1997a}
F. Julicher, A. Ajdari, and J. Prost, Reviews of Modern Physics {\bf 69},  1269
   (1997).

\bibitem{Reimann2002}
P. Reimann, Physics Reports {\bf 361},  57  (2002).

\bibitem{Astumian2002}
R.~D. Astumian and P. H\"{a}nggi, Physics Today {\bf 55},  33  (2002).

\bibitem{Hanggi2009}
P. H\"{a}nggi, Reviews of Modern Physics {\bf 81},  387  (2009).

\bibitem{Prost1994}
J. Prost, J. F. Chauwin, L. Peliti, and A. Ajdari, Physical review letters {\bf
  72},  2652  (1994).

\bibitem{Julicher1995}
F. J\"{u}licher and J. Prost, Physical Review Letters {\bf 75},  2618  (1995).

\bibitem{Julicher1997}
F. J\"{u}licher and J. Prost, Physical Review Letters {\bf 78},  4510  (1997).

\bibitem{Astumian1997}
R.~D. Astumian, Science {\bf 276},  917  (1997).

\bibitem{Rousselet1994}
J. Rousselet, L. Salome, A. Ajdari, and J. Prost, Nature {\bf 370},  446
  (1994).

\bibitem{Leibler1994}
S. Leibler, Nature {\bf 370},  412  (1994).

\bibitem{Marquet2002}
C. Marquet, A. Buguin, L. Talini, and P. Silberzan, Physical Review Letters
  {\bf 88},  168301  (2002).

\bibitem{Tierno2010}
P. Tierno, P. Reimann, T.~H. Johansen, and F. Sagu\'{e}s, Physical Review
  Letters {\bf 105},  230602  (2010).

\bibitem{Tierno2012}
P. Tierno, Physical Review Letters {\bf 109},  198304  (2012).

\bibitem{Faucheux1995}
L. P. Faucheux, L. S. Bourdieu, P. D. Kaplan, and A. J. Libchaber, Physical Review Letters
  {\bf 74},  1504  (1995).

\bibitem{Lopez2008}
B. J. Lopez, N. J. Kuwada, E. M. Craig, B. R. Long, and H. Linke, Physical Review Letters
  {\bf 101},  220601  (2008).

\bibitem{Chou1999}
C.~F. Chou, O. Bakajin, S.~W. Turner, T.~a. Duke, S.~S. Chan, E.~C. Cox, H.~G.
  Craighead, and R.~H. Austin, Proceedings of the National Academy of Sciences
  of the United States of America {\bf 96},  13762  (1999).

\bibitem{Kettner2000}
C. Kettner, P. Reimann, P. H\"{a}nggi, and F. M\"{u}ller, Physical Review E
  {\bf 61},  312  (2000).

\bibitem{Matthias2003}
S. Matthias and F. M\"{u}ller, Nature {\bf 424},  53  (2003).

\bibitem{Mennerat-Robilliard1999}
C. Mennerat-Robilliard, D. Lucas, S. Guibal, J. Tabosa, C. Jurczak, J.-Y.
  Courtois, and G. Grynberg, Physical Review Letters {\bf 82},  851  (1999).

\bibitem{Lee1999}
C.-S. Lee, B. Janko, I. Derenyi, and A.-L. Barabasi, Nature {\bf 400},  337
  (1999).

\bibitem{Olson2001}
C. J. Olson, C. Reichhardt, B. Jank\'{o}, and F. Nori, Physical Review Letters
  {\bf 87},  177002  (2001).

\bibitem{Derenyi1995}
I. Der\'{e}nyi and T. Vicsek, Physical Review Letters {\bf 75},  374  (1995).

\bibitem{Derenyi1996}
I. Der\'{e}nyi and A. Ajdari, Physical Review E {\bf 54},  R5  (1996).

\bibitem{Reimann1999}
P. Reimann, R. Kawai, C.~V. den Broeck, and P. H\"{a}nggi, Europhysics Letters
  (EPL) {\bf 45},  545  (1999).

\bibitem{Rapaport2002}
D.~C. Rapaport, Comput. Phys. Commun. 147, {\bf 147},  141  (2002).

\bibitem{Aghababaie1999}
Y. Aghababaie, G. I. Menon, and M. Plischke, Physical Review E {\bf 59},  2578
  (1999).

\bibitem{Henseler2006}
M. K\"{o}ppl, P. Henseler, A. Erbe, P. Nielaba, and P. Leiderer, Physical
  Review Letters {\bf 97},  208302  (2006).

\bibitem{Mangold2003}
K. Mangold, P. Leiderer, and C. Bechinger, Physical Review Letters {\bf 90},
  158302  (2003).

\bibitem{Chaudhuri2004}
D. Chaudhuri and S. Sengupta, Physical Review Letters {\bf 93},  115702
  (2004).

\bibitem{U.Siems2012}
U.Siems, C.Kreuter, A.Erbe, N.Schwierz, S.Sengupta, and P. P.Leiderer,
  Scientific Reports Nature Publishing Group {\bf 2},  1015  (2012).

\bibitem{Chowdhury1985}
A. Chowdhury, B.~J. Ackerson, and N.~A. Clark, Phys. Rev. Lett. {\bf 55},  833
  (1985).

\bibitem{Wei1998}
Q.-H. Wei, C. Bechinger, D. Rudhardt, and P. Leiderer, Physical Review Letters
  {\bf 81},  2606  (1998).

\bibitem{Frey1999}
E. Frey, D.~R. Nelson, and L. Radzihovsky, Phys. Rev. Lett. {\bf 83},  2977
  (1999).

\bibitem{Chaudhuri2006}
D. Chaudhuri and S. Sengupta, Physical Review E {\bf 73},  11507  (2006).

\bibitem{Frenkel2002}
D. Frenkel and B. Smit, {\em {Understanding molecular simulation: from
  algorithms to applications}} (Academic press, NY, 2002).

\bibitem{Grest1986}
G.~S. Grest and K. Kremer, Phys. Rev. A {\bf 33},  3628  (1986).

\bibitem{supmat}
See Supplemental Material for the details on equilibrium liquid-solid
  transition and density- dependent diffusivity, further characterization of
  the ratcheting dynamics in terms of frequency- and density- dependent
  effective diffusivity, and impact of {\em incommensurate} ratcheting on the
  averaged directed current.

\bibitem{Ajdari1992}
A. Ajdari and J. Prost, C R Acad Sci Paris {\bf t. 315},  1635  (1992).

\bibitem{Luczka1997}
J. Luczka, T. Czernik, and P. H\"{a}nggi, Physical Review E {\bf 56},  3968
  (1997).

\bibitem{Bao1998}
J.-d. Bao and Y.-z. Zhuo, Physics Letters A {\bf 239},  228  (1998).

\bibitem{Chaudhuri2011}
D. Chaudhuri and A. Dhar, EPL (Europhysics Letters) {\bf 94},  30006  (2011).

\bibitem{Lahtinen2001}
J. M. Lahtinen, T. Hjelt, T. Ala-Nissila, and Z. Chvoj, Physical Review E {\bf
  64},  021204  (2001).

\bibitem{Falck2004}
E. Falck and J. Lahtinen, The European Physical Journal E {\bf 13},  267
  (2004).

\end{thebibliography}

\begin{thebibliography}{1}

\bibitem{Broughton1982a}
J. Broughton, G. Gilmer, and J. Weeks, Physical Review B {\bf 25},  4651
  (1982).

\bibitem{Falck2004}
E. Falck and J. Lahtinen, The European Physical Journal E {\bf 13},  267
  (2004).

\bibitem{Lahtinen2001}
J. Lahtinen, T. Hjelt, T. Ala-Nissila, and Z. Chvoj, Physical Review E {\bf
  64},  021204  (2001).

\bibitem{Kosterlitz1973}
J.~M. Kosterlitz and D.~J. Thouless, J. Phys. C {\bf 6},  1181  (1973).

\bibitem{Nelson1979}
D.~R. Nelson and B.~I. Halperin, Phys. Rev. B {\bf 19},  2457  (1979).

\bibitem{Young1979}
A.~P. Young, Phys. Rev. B {\bf 19},  1855  (1979).

\bibitem{Sengupta2000}
S. Sengupta, P. Nielaba, and K. Binder, Physical Review. E {\bf 61},  6294
  (2000).

\bibitem{Mondal2012}
C. Mondal and S. Sengupta, Phys. Rev. E {\bf 85},  020402  (2012).

\bibitem{Chaudhuri2006}
D. Chaudhuri and S. Sengupta, Physical Review E {\bf 73},  11507  (2006).

\end{thebibliography}


\onecolumngrid
\newpage
\appendix
\section{Supplementary Information:}

\subsection{Soft-core particles -- in the absence of external potential}

\begin{figure*}[h]
\hspace{-0.7cm}
\begin{tikzpicture}
 \node[above,right] (img) at (0,0)
{\includegraphics[scale=0.82]{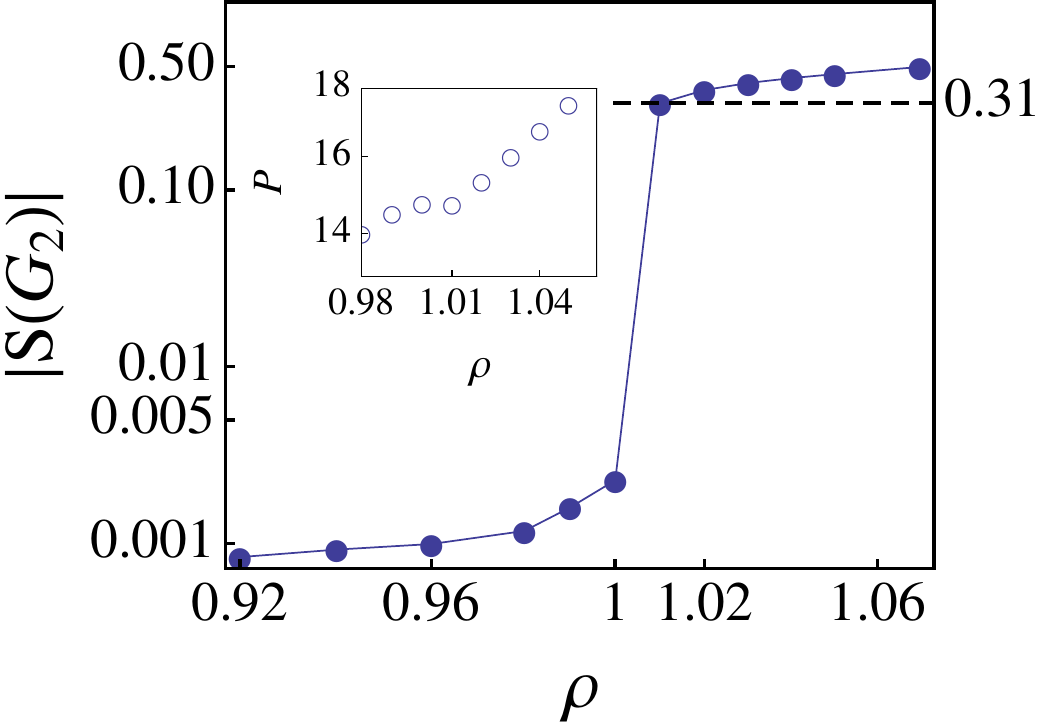}};
\node at (60pt, -32pt) {{\large \it (a)}};
\end{tikzpicture}
\hspace{0.25cm}
\begin{tikzpicture}
 \node[above,right] (img) at (0,0)
{\includegraphics[scale=0.75]{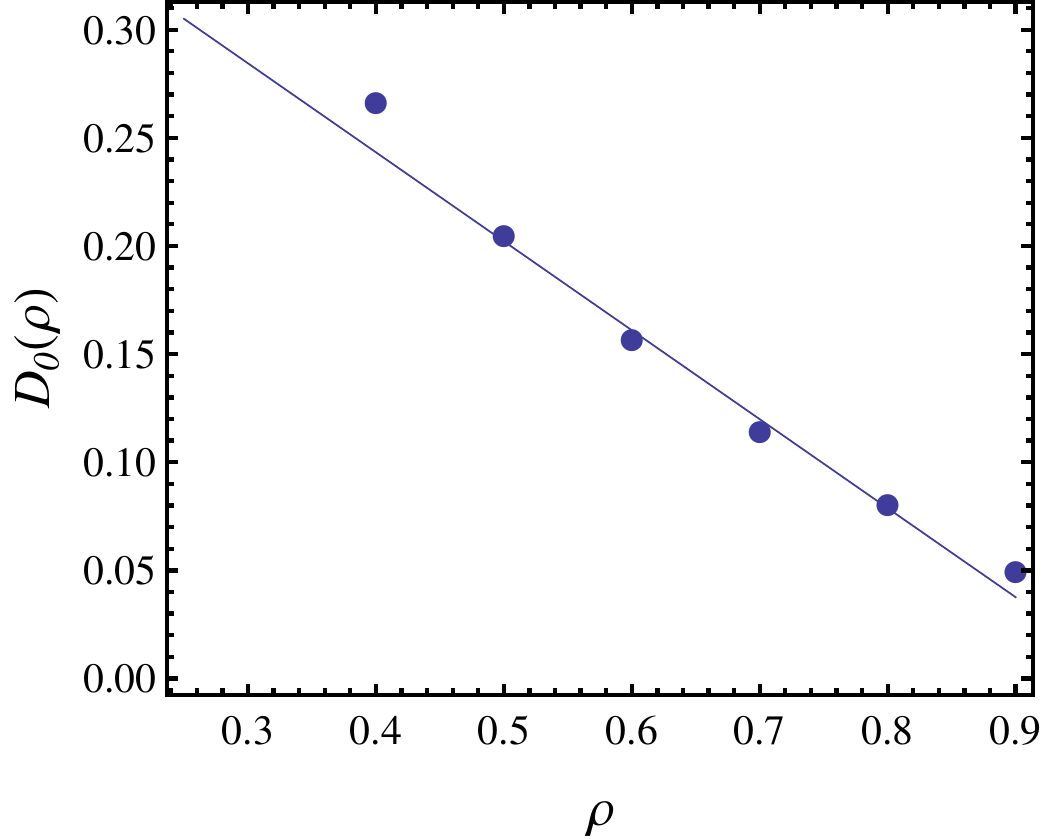}};
\node at (60pt, -32pt) {{\large \it (b)}};
\end{tikzpicture}
   \caption{($a$)~Plot of order parameter against density for
     a free system. The discontinuity of the order parameter occurs at
     a density near $\rho^\ast =1.01$. The inset
     shows pressure versus density behavior and is close to Ref.~\cite{Broughton1982a}. 
     ($b$)~Plot of the diffusion
     coefficient of the free system, averaged over both directions,
     against the density of the suspension. The solid line is a linear
     fit of $D=D_0(1-\rho/\rho_c)$ to the data with $\rho_c
     = 1.04$.}
  \label{fig:opfreesystem}
\end{figure*}

The phase behavior and dynamics of repulsively interacting colloidal suspensions
have been extensively studied in
literature~\cite{Broughton1982a,Falck2004,Lahtinen2001}.  
The two-dimensional fluid is known to undergo a freezing transition into a
triangular lattice solid, arguably via a hexatic phase, with
increasing
density~\cite{Kosterlitz1973,Nelson1979,Young1979,Sengupta2000}.
Without going into the intricacies of identifying the
hexatic~\cite{Broughton1982a,Sengupta2000}, to determine the fluid-solid transition
point we calculate the solid- order parameter $|S(G_2)|$ of the
soft-disk system, in the absence of any external driving, as a function
of the density of the system, depicted in \fref{fig:opfreesystem}(a).
At phase transition, $\rho^\ast = 1.01$,  $|S(G_2)|$ shows a discontinuous increase with
density (Fig.\ref{fig:opfreesystem}($a$) shows MD simulation result of
4096 particles). At this point the order-parameter jumps increase to a value
$|S(G_2)|=0.31$.  {In all our simulations, even in the
  presence of ratchet-driving, we identify a phase transition to a
  solid phase whenever $|S(G_2)|$ crosses the value $0.31$.}  Note
that this value is close to the phase transition criterion of
$|S(G_2)|=0.35$ used in recent literature~\cite{Mondal2012}.  As depicted 
in the inset of Fig.\ref{fig:opfreesystem}($a$), the pressure versus density behavior 
shows non-monotonicity signifying  a phase-coexistence of a 
solid with density $\rho_s =1.01$ and liquid with $\rho_l=0.99$~\cite{Broughton1982a}.

In the overdamped regime, the internal relaxation is due to diffusive
motion of particles.
The density dependence of diffusivity $D(\rho)$ in soft-disk particles
has been previously studied by Lahtinen {\it et. al}
\cite{Falck2004,Lahtinen2001} and exhibits a linear dependence on
$\rho$. Using our MD simulations, we studied the same to obtain the
linear density dependence $D=D_0(1-\rho/\rho_c)$, where $D_0$ is the
diffusivity of a noninteracting system, and the fitted value $\rho_c =
1.04 $ (Fig.\ref{fig:opfreesystem}($b$)).

\subsection{Soft-core particles -- under stochastic ratcheting}
In order to study the effect of stochastic ratcheting, the system was
evolved under the potential $\be U_{\rm ext}(y,t)=\be V_0(t)
\left[\sin\left( 2 \pi y/\l \right) +\alpha \sin\left(4\pi y/\l
  \right) \right]$, with $\be V_0(t)$ switching between 0 and $1$, and
asymmetry parameter $\a=0.2$.
The switching is done stochastically with a rate $f$. During our MD
simulation, at each time step the value of the external potential
$V(t)$ is switched between 0 and 1 with a probability $f\d t$, and
left unaltered with a probability $(1-f\d t)$, where $\d t$ is the
size of integration time-step. We have always waited for the system to
reach steady state before collecting data presented in all our
analysis.

\begin{figure*}[t]
\begin{tikzpicture}
 \node[above,right] (img) at (0,0)
{\includegraphics[scale=0.7]{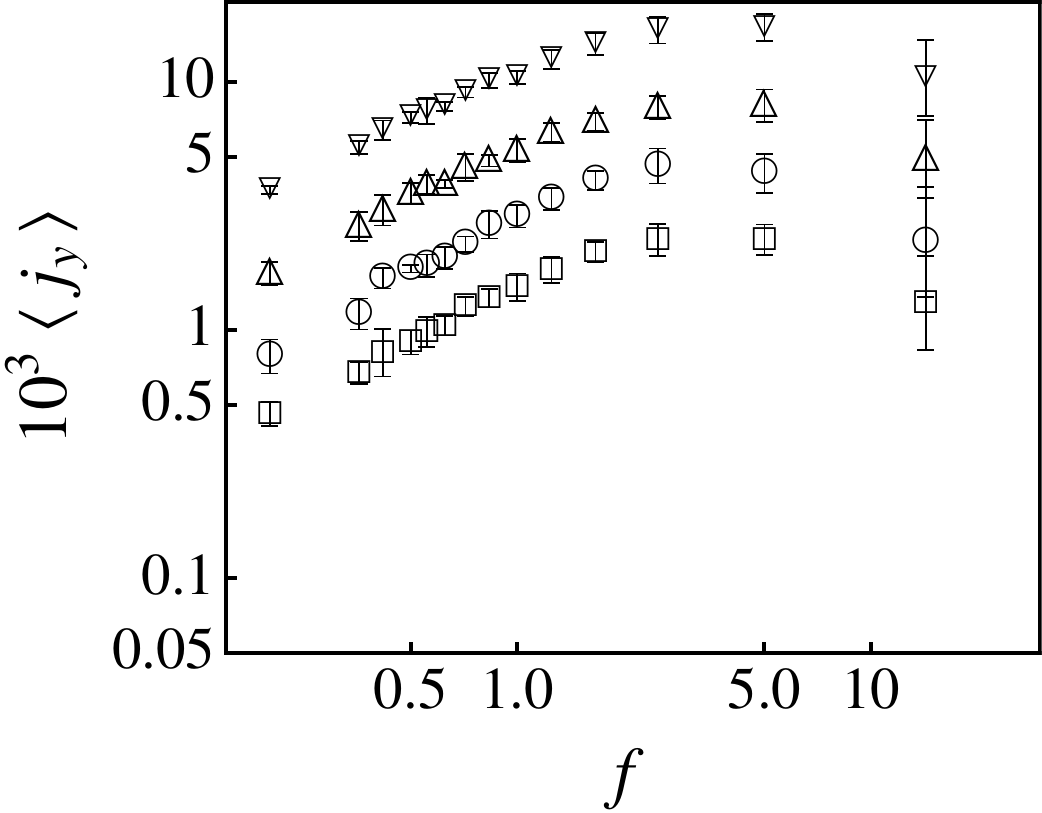}};
\node at (10pt, 70pt) {{\large \it (a)}};
\end{tikzpicture}
\hspace{1.5cm}
\begin{tikzpicture}
 \node[above,right] (img) at (0,0)
{\includegraphics[scale=0.7]{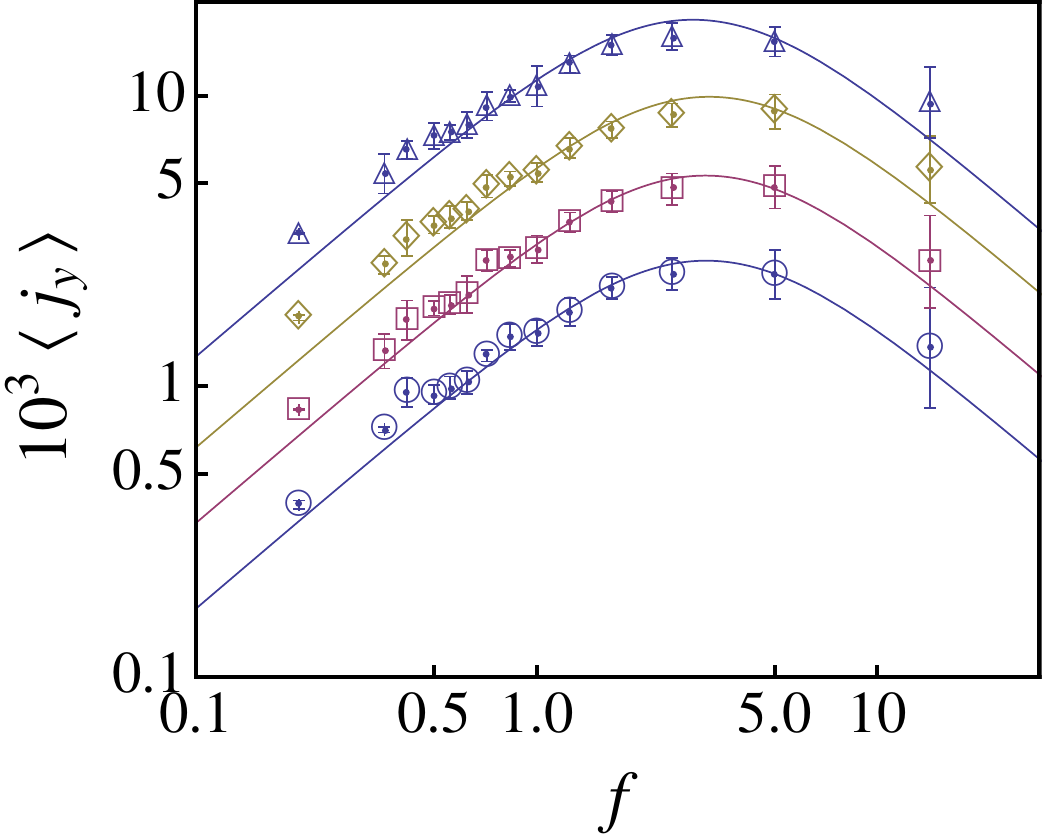}};
\node at (10pt, 70pt) {{\large \it (b)}};
\end{tikzpicture}
\begin{tikzpicture}
 \node[above,right] (img) at (0,0)
{\includegraphics[scale=0.7]{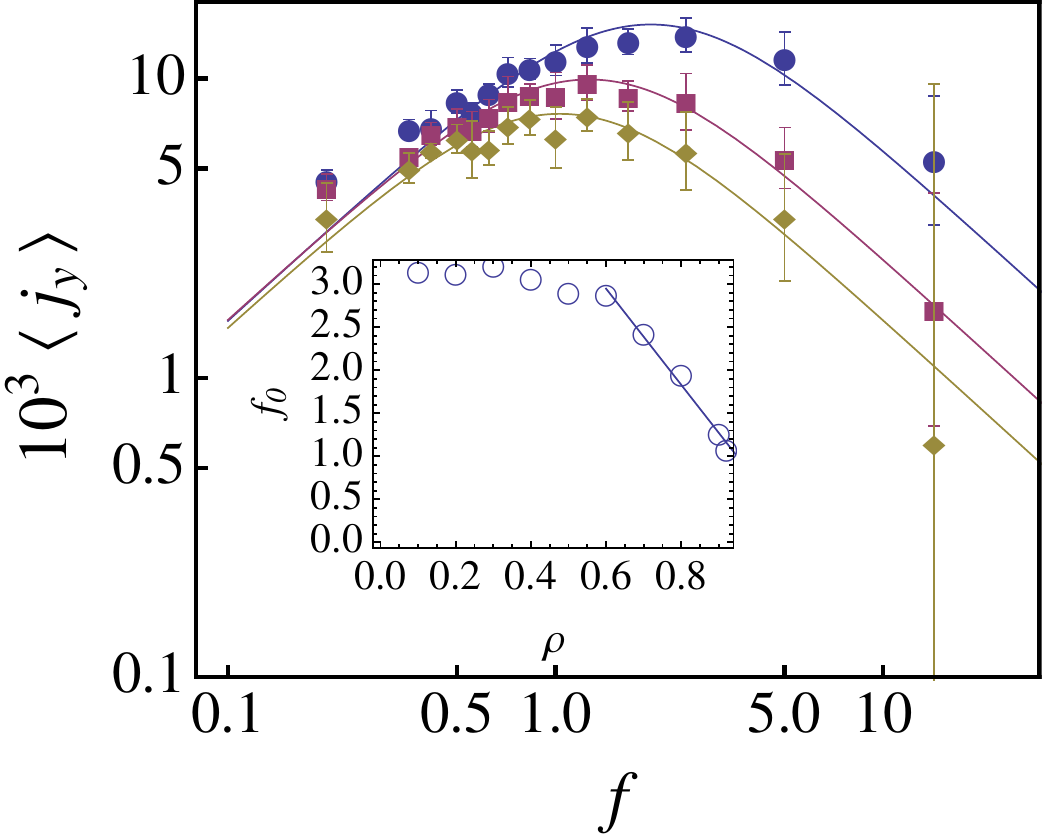}};
\node at (10pt, 70pt) {{\large \it (c)}};
\end{tikzpicture}
\hspace{1.5cm}
\begin{tikzpicture}
 \node[above,right] (img) at (0,0)
{\includegraphics[scale=0.52]{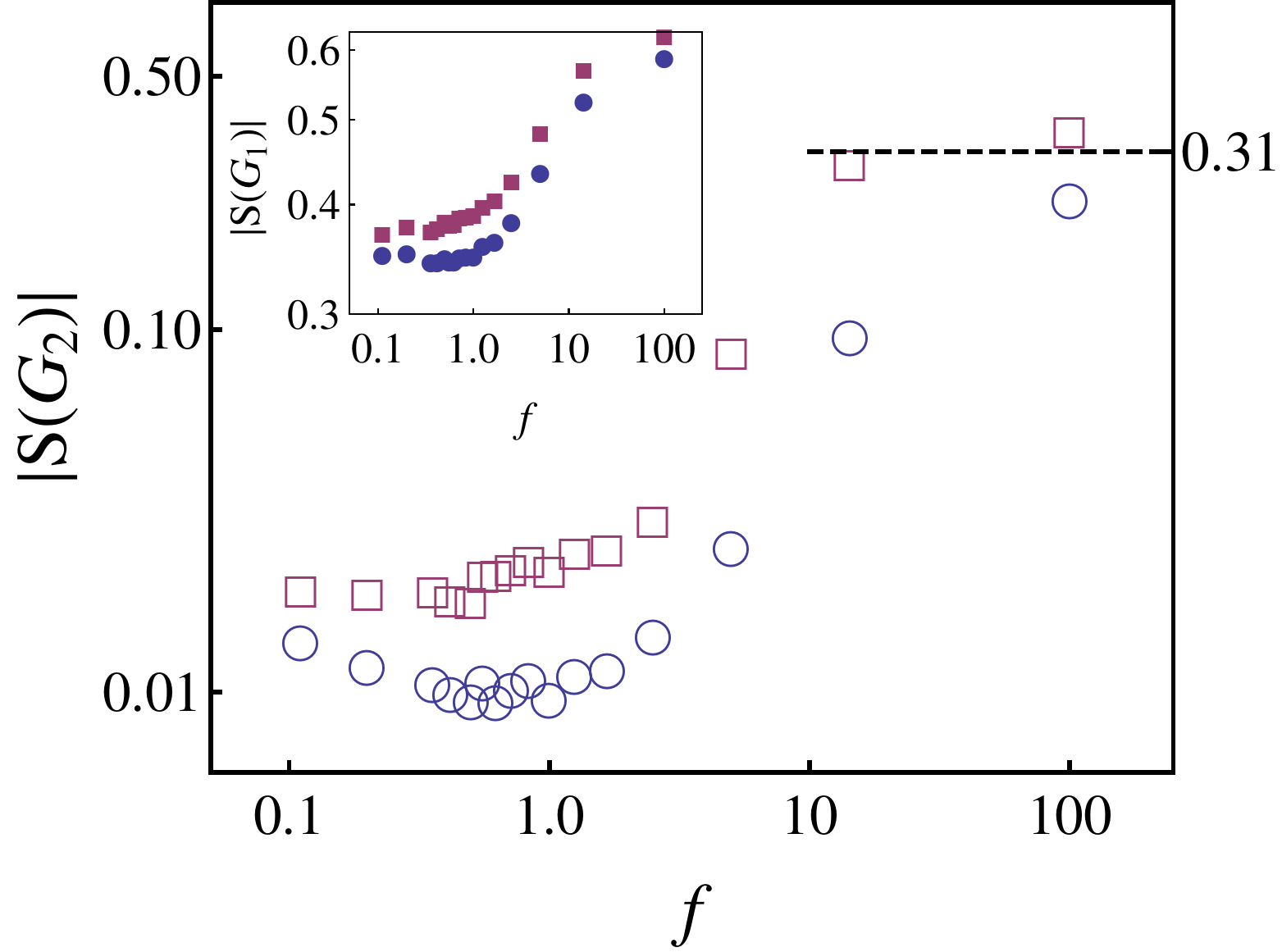}};
\node at (10pt, 70pt) {{\large \it (d)}};
\end{tikzpicture}

\hspace{0.5cm}
\caption{(Color Online) ($a$)~Plot of the averaged directed current
  for a non-interacting system driven by a periodic potential with
  periodicity $1\s$, for densities $\rho =0.1$ ({\Large $\circ$}), $0.2$
  ({\large $\square$}), $0.3$ ($\bdiamond$) and $0.6$
  ($\triangle$). The resonance in the measured flux occurs at a
  fixed frequency $f_0 = 3.5 $.  
  ($b$)~Plot of
  the measured flux for an interacting system driven by a periodic
  potential with peroidicity $1\s$, for densities $\rho =0.1$
  ({\color{myblue} \large $\circ$}), $0.2$ ({\color{mypurple}
    $\square$}), $0.3$ ({\color{myokker} $\bdiamond$}) and $0.6$
  ({\color{myblue} $\triangle$}). The solid lines are fit to the data
  using the functional form of Eq.~(2) in the main text.
  ($c$)~Plot of the measured flux for an interacting
  system driven by a periodic potential with peroidicity $1\s$, for
  densities $\rho =0.8$ ({\color{myblue} \Large $\bullet$}), $0.92$
  ({\color{mypurple} $\blacksquare$}) and $0.98$ ({\color{myokker}
    $\fdiamond$}). The solid
  lines are fit to the data using the functional form of Eq.~(2) in
  the main text. The inset depicts the dependence of the resonance
  frequency on the density of the suspension. The solid line is a
  linear fit of $f_0 \sim (1-\rho/\rho_c)$ to the data with
  $\rho_c\approx 1.13$.
($d$)~Plot of the measured order parameter for the reciprocal lattice vectors
  $\mathbf{G}_2$ (empty symbols) and $\mathbf{G}_1$ (inset, filled symbols)
  for densities $\rho = 0.94$~({\color{myblue} \Large $\bullet$,
    $\circ$}) and $0.96$~({\color{mypurple}\footnotesize
    $\blacksquare$,$\square$}).}
  \label{fig:flux}
\end{figure*}
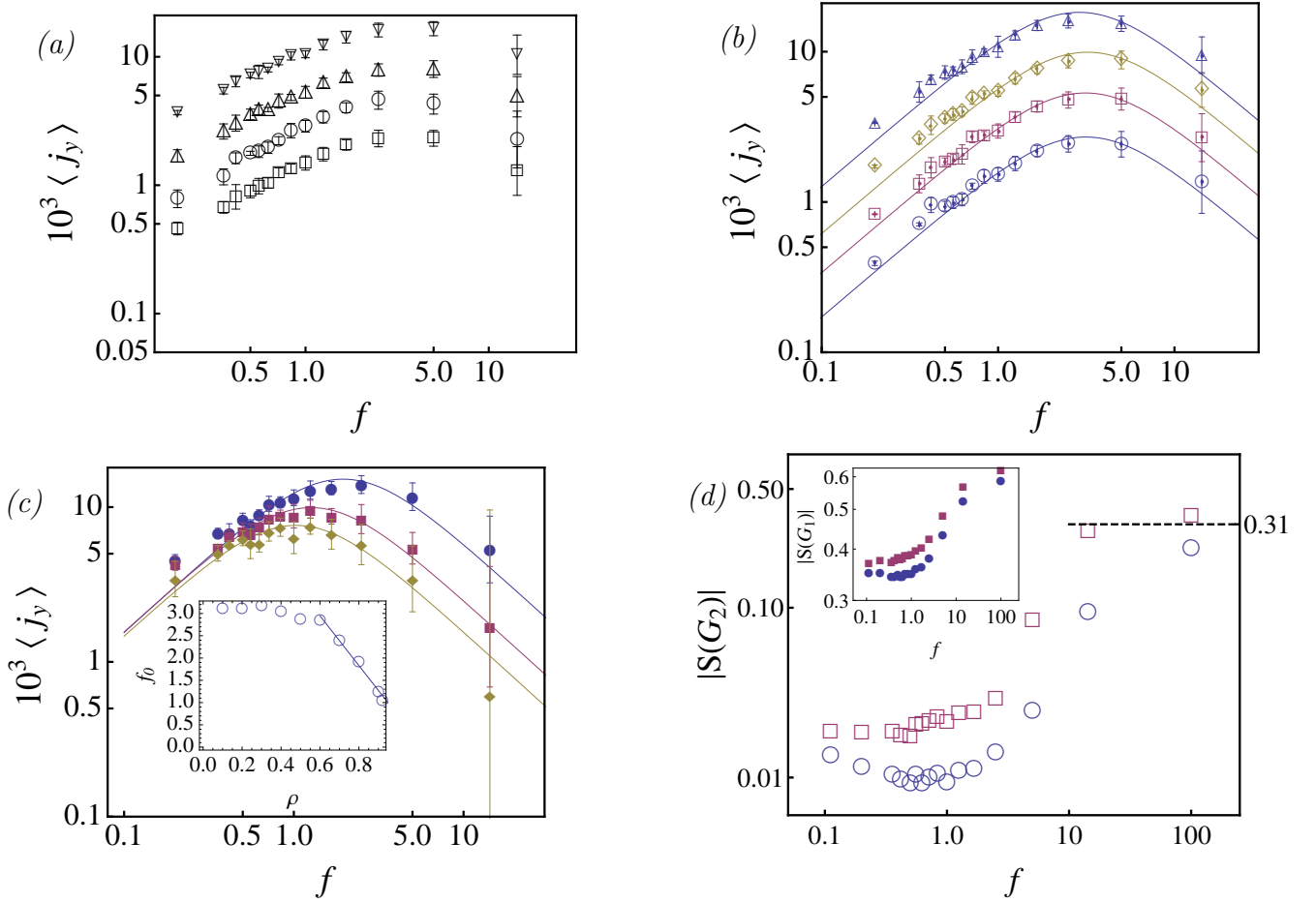

In this section, we first focus on systems driven by a ratchet of
fixed periodicity $\l=1 \s$.  The driving potential is not
commensurate with the density, unlike the system discussed in the main
text. The impact on the non-interacting system itself is different, as
$\l$ is no more a function of density.
%
The integrated directed current $\la j_y \ra$ exhibits 
resonance at a fixed value of the frequency of $f_0 = 3.5 $,
independent of densities (see Fig.~\ref{fig:flux}~($a$)). The
ballistic time-scale $\t_b$ to travel a potential-valley of length $\l$ 
follows the kinematic relation $\l\sim (U_0/\l) \t_b^2$ leading to
$\t_b \sim \l/\sqrt{U_0}$, independent of density as $\l$ itself is a constant. 
As a result the resonance frequency is also constant.

If, instead, an interacting system is driven by the same ratcheting
potential with $\l=1\s$, the resonance frequency remains approximately
constant at lower densities ($\rho \lesssim 0.5$) indicating a
free-particle like ballistic motion.  However, at higher densities the
resonance frequency drops { approximately linearly with density
  (inset of Fig.~\ref{fig:flux}~($c$))}. This indicates a cross-over
from ballistic to diffusive transport with diffusive time-scale $\t_D
\sim \l^2/D_0(1-\rho/\rho_c)$, and the corresponding resonance
frequency $\sim (1-\rho/\rho_c)$. This behavior should be contrasted
against the $f \sim \rho(1-\rho/\rho_c)$ behavior for {\em
  commensurately } ratcheting soft- disks at high densities (Fig.3 of
main text).  Note that the reasoning used here is similar to that we
applied for interacting particles driven by a commensurate ratchet
(See Eq.s (2)--(4) of main text).

It is known that, in case of laser induced freezing (LIF), the
soft-core system of particles undergo a modulated liquid to solid
transition near $\rho=0.95$ and a periodic commensurate potential of
strength $\be U_0=1$~\cite{Chaudhuri2006}. As we argued in the main
text, at very high frequencies of weakly asymmetric commensurate
ratcheting, one expects to recover fluid-solid transitions similar to
LIF.  In Fig.~\ref{fig:flux}~($d$) we present the dependence on the
ratcheting frequency $f$ of the solid order parameter
$S(\mathbf{G}_2)$ (defined in main text) at densities $\rho =0.94$ and
$0.96$. This shows that at high enough frequencies $f \gg 10$,
although the effective time-integrated periodic potential strength
$\be U_0 <1$, the system at $\rho=0.96$ already shows freezing,
a reminiscent of LIF.

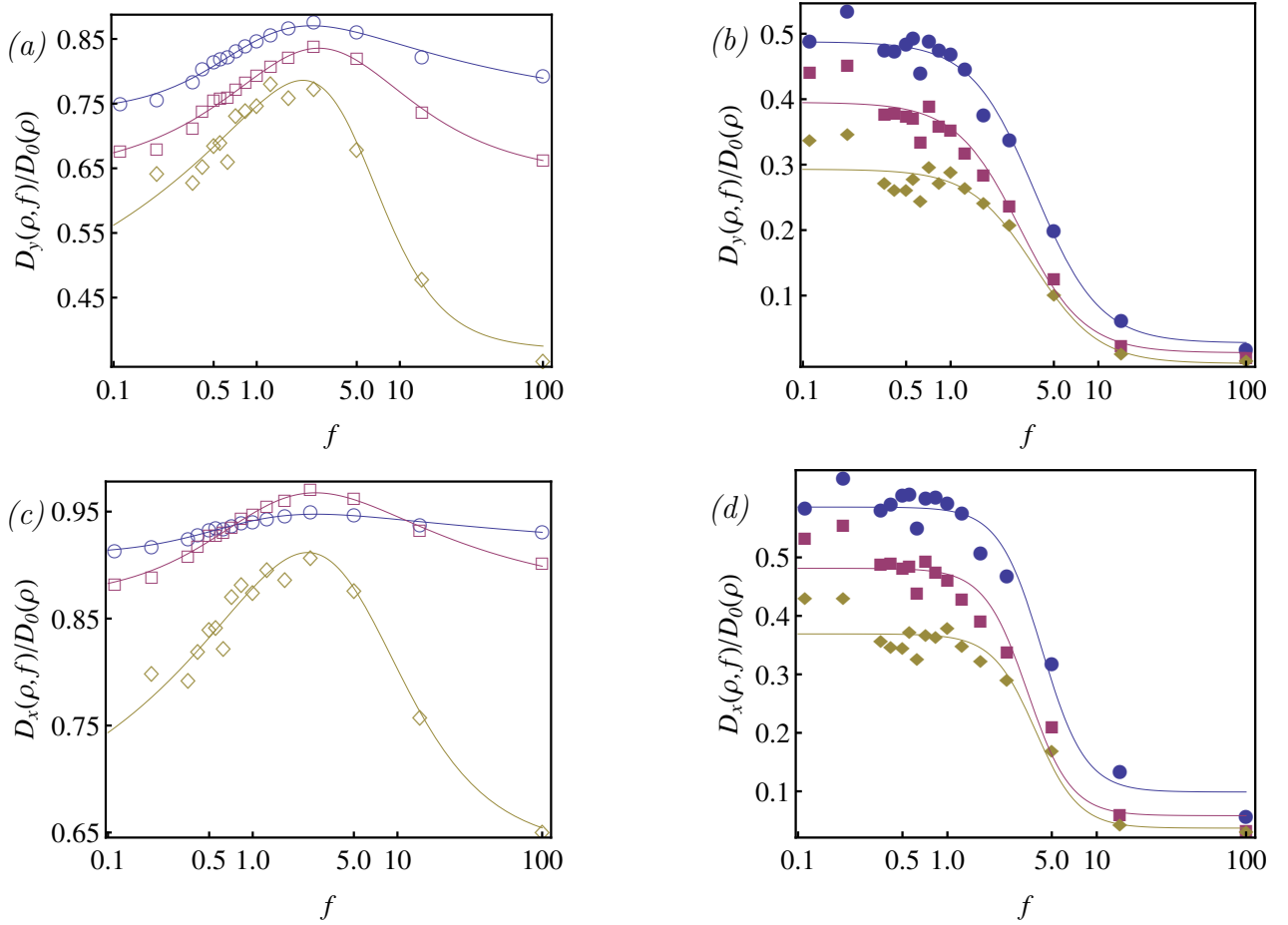
\begin{figure*}[t]
\begin{tikzpicture}
 \node[above,right] (img) at (0,0)
{\includegraphics[scale=0.7]{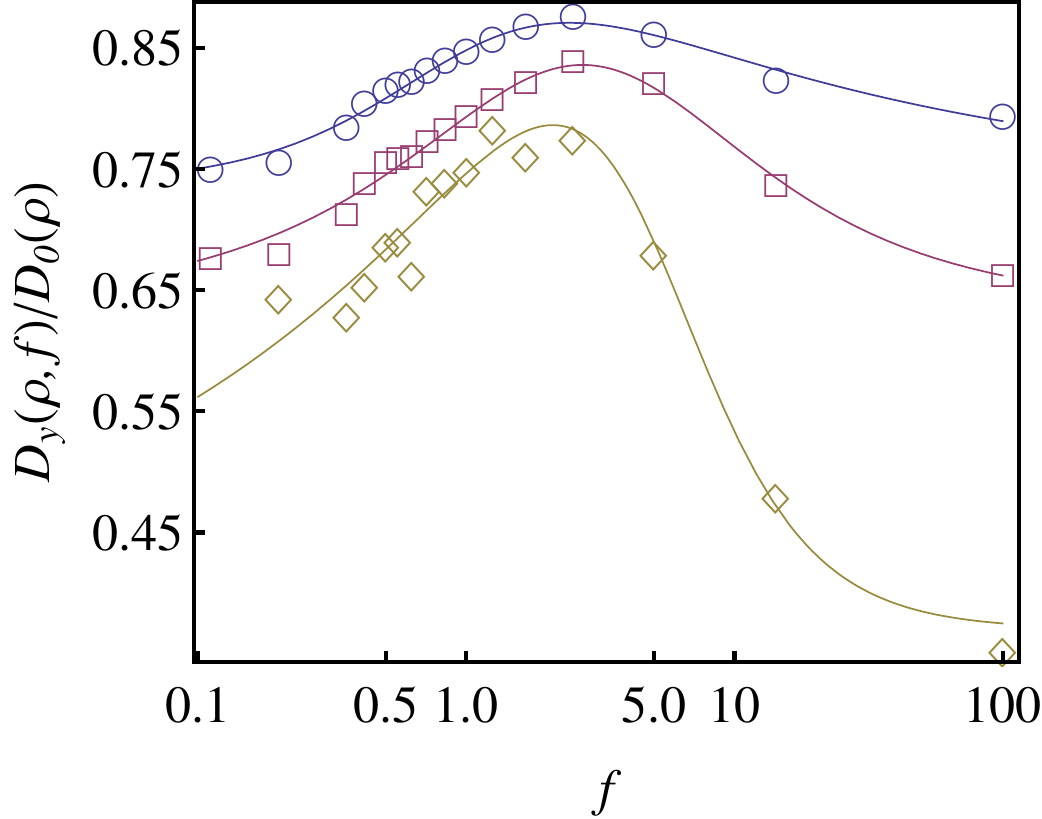}};
\node at (10pt, 70pt) {{\large \it (a)}};
\end{tikzpicture}
\hspace{1.5cm}
\begin{tikzpicture}
 \node[above,right] (img) at (0,0)
{\includegraphics[scale=0.7]{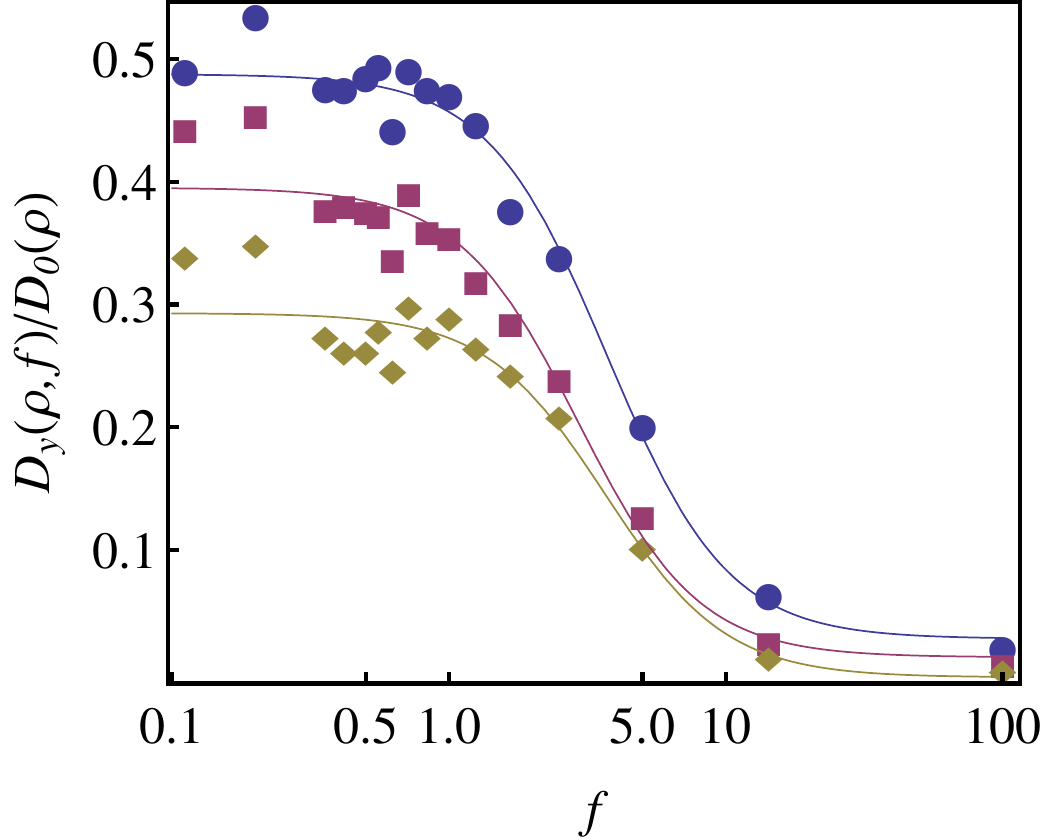}};
\node at (10pt, 70pt) {{\large \it (b)}};
\end{tikzpicture}
\begin{tikzpicture}
 \node[above,right] (img) at (0,0)
{\includegraphics[scale=0.7]{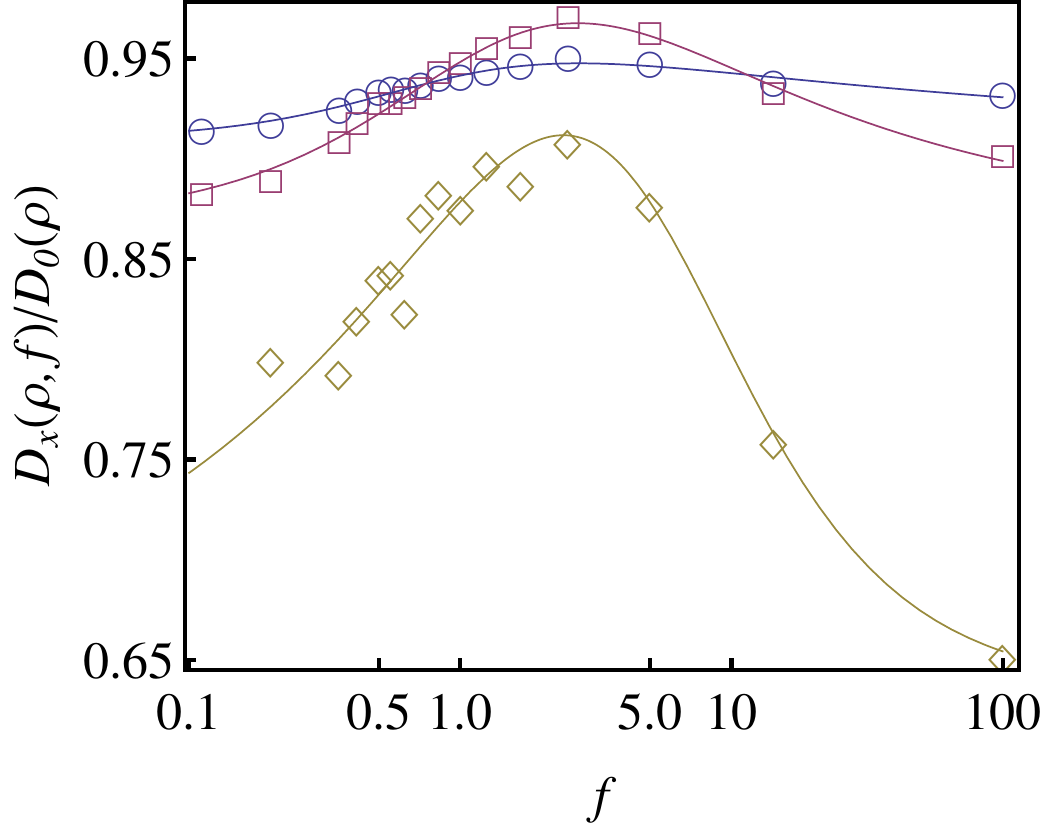}};
\node at (10pt, 70pt) {{\large \it (c)}};
\end{tikzpicture}
\hspace{1.5cm}
\begin{tikzpicture}
 \node[above,right] (img) at (0,0)
{\includegraphics[scale=0.7]{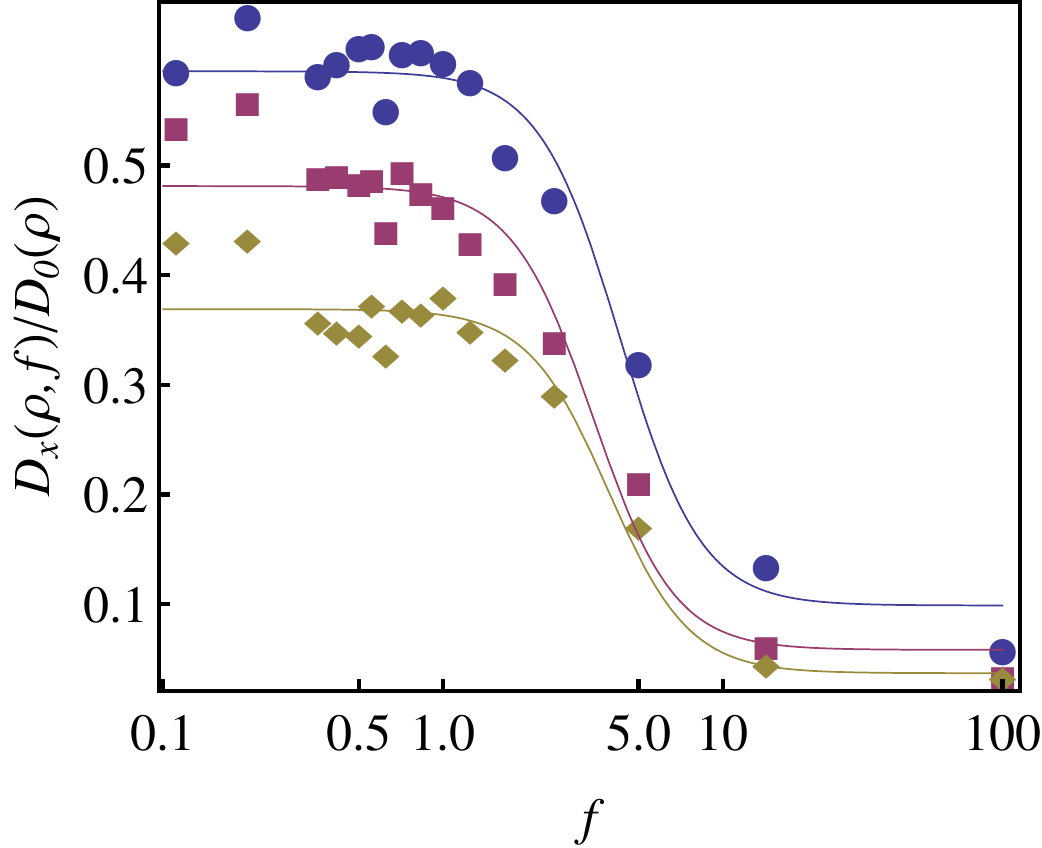}};
\node at (10pt, 70pt) {{\large \it (d)}};
\end{tikzpicture}
\caption{Plot of the measured diffusion coefficient,normalized by the
  diffusion coefficient of a free system $D_0(\rho)$, for an
  interacting system driven by a periodic and a commensurate potenial,
  along (in ($a$) and ($b$)) and perpendicular
  (in ($c$) and ($d$)) to the direction of the
  external drive , as a function of frequency for densities of the
  suspension $\rho =0.2$ ({\color{myblue} {\Large $\circ$}}),
  $0.5$ ({\color{mypurple} $\square$}), $0.8$ ({\color{myokker}
    {$\bdiamond$}}), $0.92$ ({\color{myblue} {\Large $\bullet$}}),
  $0.94$ ({\color{mypurple} $\blacksquare$}) and $0.96$
  ({\color{myokker} {$\fdiamond$}}). The solid lines are guide to the
  eye. }
\label{fig:diff_coeff}
\end{figure*}

\begin{figure*}[t]
\begin{tikzpicture}
 \node[above,right] (img) at (0,0)
{\includegraphics[scale=0.7]{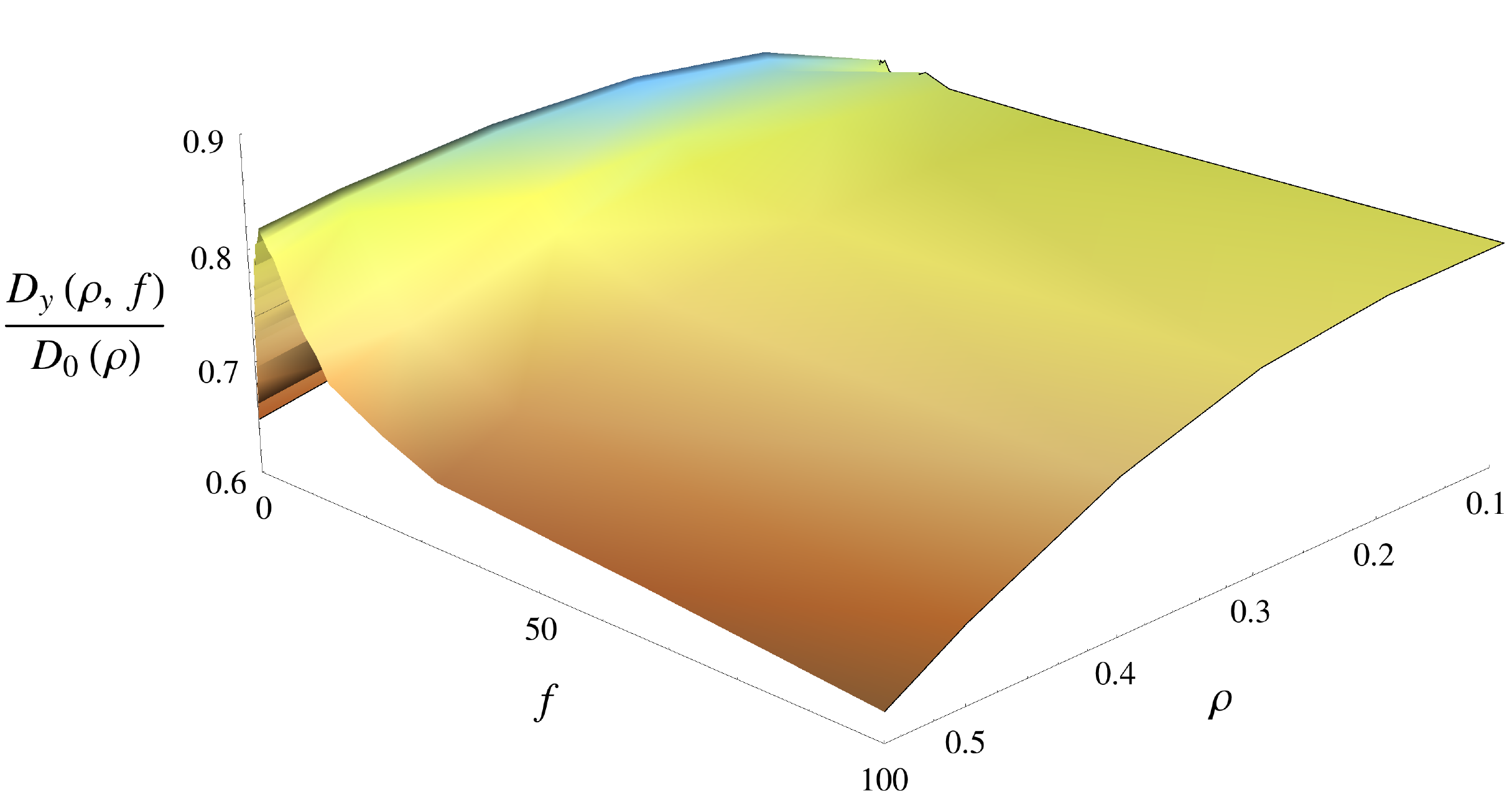}};
\node at (10pt, 70pt) {{\large \it (a)}};
\end{tikzpicture}
\hspace{1.5cm}
\begin{tikzpicture}
 \node[above,right] (img) at (0,0)
{\includegraphics[scale=0.7]{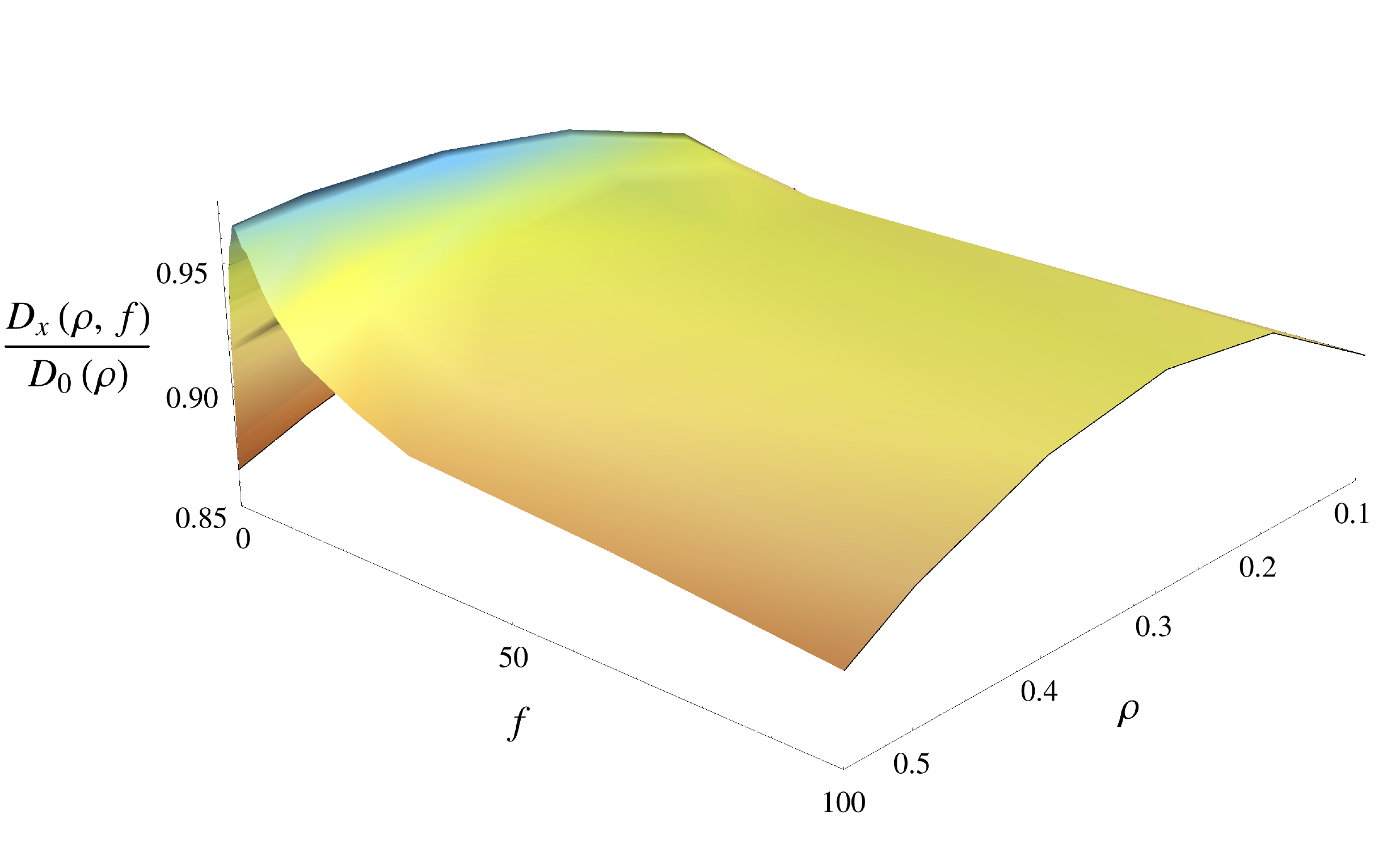}};
\node at (10pt, 70pt) {{\large \it (b)}};
\end{tikzpicture}
\caption{Plot of the measured diffusion coefficient,normalized by the
  diffusion coefficient of a free system $D_0(\rho)$, for an
  interacting system driven by a periodic and a commensurate potential,
  along ($a$) and perpendicular
  ($b$) to the direction of the
  external drive, as a function of frequency and density.
  }
\label{fig:diff_coeff_3d}
\end{figure*}

Finally, we consider the asymmetric diffusivity of the soft-core
particle system, driven by a ratchet with periodicity commensurate
with the density. We obtain the density and ratcheting-frequency
dependence of the diffusivity.  In Fig.~\ref{fig:diff_coeff}, we show
the frequency dependence of diffusion coefficients $D_y$, $D_x$ along
and perpendicular to the direction of ratcheting drive, respectively.
We observe that $D_y < D_x$, since to diffuse in the direction of
driving, particles have to climb potential barriers.  As with the
directed current, we also observe a resonance in the diffusion
coefficient when the density is relatively small (see
\fref{fig:diff_coeff}(a) and (c)).
The properties of $D_{x,y}(f,\rho)$ is further presented as a surface
plot in Fig.~\ref{fig:diff_coeff_3d}, in the low density regime where
the resonance structure is seen. The magnitude of both $D_x$ and $D_y$
increases with density, the effect being more pronounced for $D_x$ and
persists for densities lower than $\approx 0.3$ (see
Fig.~\ref{fig:diff_coeff_3d}), beyond which the magnitude of the
diffusion constants steadily decreases.
At higher densities, the resonance structure is lost, the
diffusivities remain approximately constant at low frequencies,
decreasing at frequencies $f \gtrsim1$ (Fig.s~\ref{fig:diff_coeff}($b$)
and (d)).

\end{document}